\documentclass{aa}

\usepackage{graphicx} \usepackage{amssymb, amsmath}

\begin{document}

\title{Cosmic shear from STIS pure parallels\thanks{Based on
observations made with the NASA/ESA \emph{Hubble Space Telescope},
obtained at the Space Telescope Science Institute (STScI) and from the
data archive at the STScI, which is operated by AURA, Inc., under NASA
contract NAS 5-26555}} 
\subtitle{II. Analysis}

\author{H.~H{\"a}mmerle\inst{1,2} 
 \and J.-M.~Miralles\inst{1,3} 
 \and P.~Schneider\inst{1,2} 
 \and T.~Erben\inst{2,4,5} 
 \and R.~A.~E.~Fosbury\inst{3} 
 \and W.~Freudling\inst{3} 
 \and N.~Pirzkal\inst{3} 
 \and B.~Jain\inst{6} 
 \and S.~D.~M.~White\inst{2} } 

\institute{ Institut f{\"u}r Astrophysik und Extraterrestrische
 Forschung der Universit{\"a}t Bonn, Auf dem H{\"u}gel 71, D-53121
 Bonn, Germany 
 \and Max-Planck-Institut f{\"u}r Astrophysik, Karl-Schwarzschild
 Str.\ 1, D-85748 Garching, Germany 
 \and ST-ECF, Karl-Schwarzschild Str.\ 2,
 D-85748 Garching, Germany
 \and Institut d'Astrophysique de Paris, 98bis Boulevard Arago,
 F-75014 Paris, France 
 \and Observatoire de Paris, DEMIRM 61, Avenue de l'Observatoire,
 F-75014 Paris, France 
 \and Departement of Physics and Astronomy, University of Pennsylvania
 209 S.\ 33rd Street, Philadelphia, PA 19104 USA 
}

\offprints{H.~H{\"a}mmerle, \email{hanne@astro.uni-bonn.de}}

\abstract{We report on the marginal detection of cosmic shear on sub-arcmin
scales with archive data from the STIS camera on board HST. For the
measurement 121 galaxy fields with a field of view of $51'' \times
51''$ are used to obtain an rms cosmic shear of $\sim 4\%$ with
$1.5\sigma$ significance. This value is consistent with
groundbased results obtained by other groups on larger scales, and
with theoretical predictions for a standard $\Lambda$CDM cosmology. To
show the suitability of STIS for weak 
shear measurements  we carefully investigated the stability of the
PSF. We demonstrate that small temporal changes do not affect the
cosmic shear measurement by more than $\sim 10\%$. We also discuss the
influence of various weighting and selection schemes for the galaxy
ellipticities.  
}

\date{Received 10 Ocotber 2001 / Accepted 31 January 2002}  
\titlerunning{Cosmic shear from STIS pure parallels} 
\authorrunning{H.\ H{\"a}mmerle et al.}  
\maketitle

\def\ave#1{\left\langle #1 \right\rangle} 
\def\e{e}
\def\rh{r_\mathrm{h}} \def\abs#1{\left| #1 \right|} 
\def\arcsecf{\hbox{$.\!\!^{\prime\prime}$}} 
\def\arcminf{\hbox{$.\!\!^{\prime}$}}
\def\Pg{P^\gamma} 
\def\cs{\ave{\overline{\gamma}^2} }
\def\csn{\overline{\gamma}_n^2 }
\def\sigcs{\sigma_{\ave{\overline{\gamma}^2}} }

\begin{section}{Introduction}
The tidal gravitational field due to the inhomogeneous distribution
of matter in the Universe causes the distortion of light bundles from
distant sources. As a consequence, the observed images of these sources are
deformed. This tidal distortion yields a small but observable imprint
on the distribution of galaxy ellipticities, an effect called `cosmic
shear'. As pointed out by \cite{Bea91}~(1991), \cite{ME91}~(1991) and
\cite{K92}~(1992), the statistical properties of the shear 
field reflect the statistical properties of the (dark) matter
distribution in the Universe. Refined theoretical predictions,
accounting for the non-linear evolution of the large-scale structure
(e.g.\ \cite{JS97}~1997; \cite{Bea97}~1997; \cite{K98}~1998;
\cite{Schneider98a}~1998a) and numerical simulations based on N-body
results for cosmic structure evolution (e.g., \cite{vWea99}~1999,
\cite{JSW00}~2000) yield expected shear amplitudes of about one 
percent on scales of a few arcminutes, depending on cosmological
parameters and, in particular, on the normalization of the dark matter
power spectrum (for recent reviews on the topic, see \cite{M99}~1999;
\cite{BS01}~2001).

Because of the small scale of this effect, investigating it
observationally is challenging. Early attempts
(e.g.\ \cite{Mea94}~1994; \cite{Schneider98b}~1998b) where plagued by
the small data sets. Furthermore, dedicated image analysis
software was needed to correct for observational effects like
atmospheric seeing, anisotropic PSFs etc.
(e.g, \cite{BM95}~1995; Kaiser et al.~1995 (hereafter
referred to as \cite{KSB}); \cite{LK97}~1997).

In March 2000, four groups nearly simultaneously announced the
detection of cosmic shear on angular scales between $\sim 1\arcmin$ and
$\sim 10\arcmin$ (\cite{Bea00}~2000; \cite{KWL00}~2000;
\cite{vWea00}~2000; \cite{Wea00}~2000). All four groups presented
results from groundbased wide-field imaging
observations, at the level expected from
currently popular cosmological models; particularly noteworthy is the
impressive agreement between the four independent results. 
\cite{Mea01}~(2001) measured cosmic shear on an angular scale of a
few arcminutes, using a survey carried out with the VLT, and again
found results in agreement with the others. \cite{vWea01}~(2001)
presented the currently most accurate cosmic shear measurement, 
based on about 6.5 square degrees of CFHT wide-field imaging
data. With the precision of these measurements, interesting
cosmological constraints can be obtained from these data. 

To extend these results to smaller angular scales where the
non-linear effects of structure formation are more pronounced, other
observing strategies are preferable. The noise of cosmic shear
measurements is due mainly to two sources: intrinsic ellipticity of
the source galaxies, and sampling variance. The relative importance of
these two noise terms depends on the angular scale (e.g.,
\cite{K98}~1998); for small-scale cosmic shear measurements, the
intrinsic 
ellipticity becomes dominant. To reduce this noise contribution, one
should strive for imaging data with a large number density of galaxies,
i.e., very deep images.

On the other hand, fainter galaxies have a  smaller  angular
extent. The observed images are affected by a PSF smearing which must
be corrected. The smaller galaxies need larger corrections and,
accordingly, the corresponding uncertainty is larger. This means
that shear measurements with groundbased telescopes have a natural
limit, taking deeper images will not provide a significant increase
in the number density of galaxies which can be used for a shear
analysis. To measure accurate shapes of very faint and small
galaxy images, space-based observations are required.

The Hubble Space Telescope currently carries two optical cameras,
WFPC2  and STIS. Data from both of these instruments can in
principle be used to study cosmic shear. The advantage of WFPC2 is its
larger field-of-view, whereas the STIS imager has higher throughput
and smaller pixel size, which is better adapted to the diffraction
limit of the 
HST. In addition, it is known from weak lensing studies of clusters
with WFPC2 (e.g., \cite{Hea98}~1998) that the PSF shows relatively
large anisotropies towards the edges of the three WFC-chips. In
contrast to wide-field groundbased images, where the PSF anisotropy
is measured from the stars on the same image for which the shear
analysis is performed, such a procedure is impossible for HST
images: the fields are too small to detect a sufficient number of stars 
to model the PSF anisotropy. Since there are also
indications that the PSF in WFPC2 changes with time
(\cite{Hea98}~1998), we dismissed the idea of cosmic shear
measurements with WFPC2 
(but see \cite{Rea01}~2001).  Instead, analysing early STIS images,
we concluded that the PSF anisotropy was very small, and did not
change significantly with time.

With the initiation of a public parallel program with STIS in June
1997, we decided to use these archival data for a cosmic shear
study. The data set that we have analysed has been described in
Pirzkal et al.~(2001; hereafter \cite{PaperI}); briefly, we have produced a
set of 498 coadded frames, some of which contain many
stars and thus can be used to investigate the PSF, whereas others are
dominated by galaxies and can be used for the shear analysis (see
\cite{PaperI} and Sect.~\ref{sc:cat} below for details). 

The rest of the paper is organized as follows: In Sect.~\ref{sc:cat} we briefly
describe our data set, the field selection and the data analysis. We
devote Sect.~\ref{sc:PSF} to a study of the PSF anisotropy, using images
containing many stars, and demonstrate that the variations of the PSF
anisotropy in time are indeed very small. A description of the cosmic
shear analysis is given in Sect.~\ref{sc:shear}, including our
procedure for 
correcting for PSF anisotropy and smearing. We derive
the maximum variations of the resulting mean shear per galaxy field
when using the PSFs obtained from different star fields, and
conclude that the STIS PSF anisotropy will not affect a cosmic shear
measurement above the $\sim 10\%$ level. Results are presented
in Sect.~\ref{sc:res}. We obtain a $\sim 1.5\sigma$ detection of the
rms shear  on the scale of the STIS fields, with
$\sqrt{\cs}=3.87^{+1.29}_{-2.04}\%$.  This value extends
previous cosmic shear measurements to smaller 
angular scales. The influence of various weighting and selection
schemes for the galaxy ellipticities is studied; in
particular, we show that using different PSF anisotropy functions
for the ellipticity corrections leaves the result basically
unchanged. We tested our scheme with simulations and verified that it
is able to recover a cosmic shear variance to within $\sim 15\%$ of
the input value. We conclude in Sect.~\ref{sc:diss} with a discussion of our
results and an outlook on the currently running Pure Parallel GO 
Program with STIS.

\end{section}

\begin{section}{Catalogue production and field selection}\label{sc:cat}
A preselection of the 498 associations defined in \cite{PaperI} was
done based on a visual inspection of the fields. We rejected fields in
which the individual exposures 
were unsuitable for our project; reasons to not use a field were: 
bright stars and/or  too many stars with  diffraction spikes, (almost)
empty fields (less than 5 objects per field), lower
image quality than the rest of the images,  or failure to converge of
the  iterative co-addition procedure  described in \cite{PaperI}. Some
examples of fields which were rejected are 
shown in Fig.~\ref{fig:notused}.

\begin{figure*}                             
 \centering
  \begin{tabular}{ccc}
  \includegraphics[width=4cm]{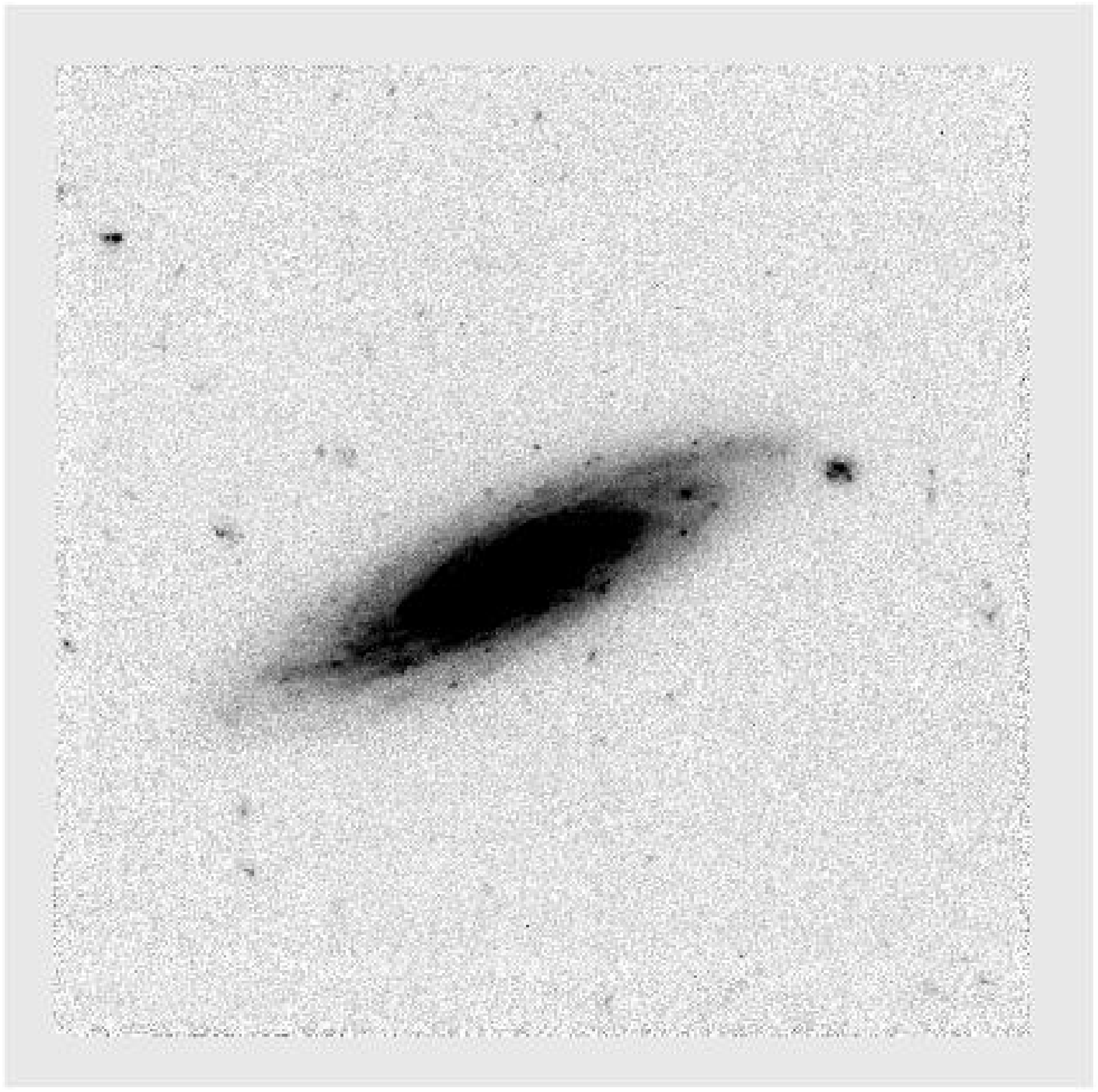} &
  \includegraphics[width=4cm]{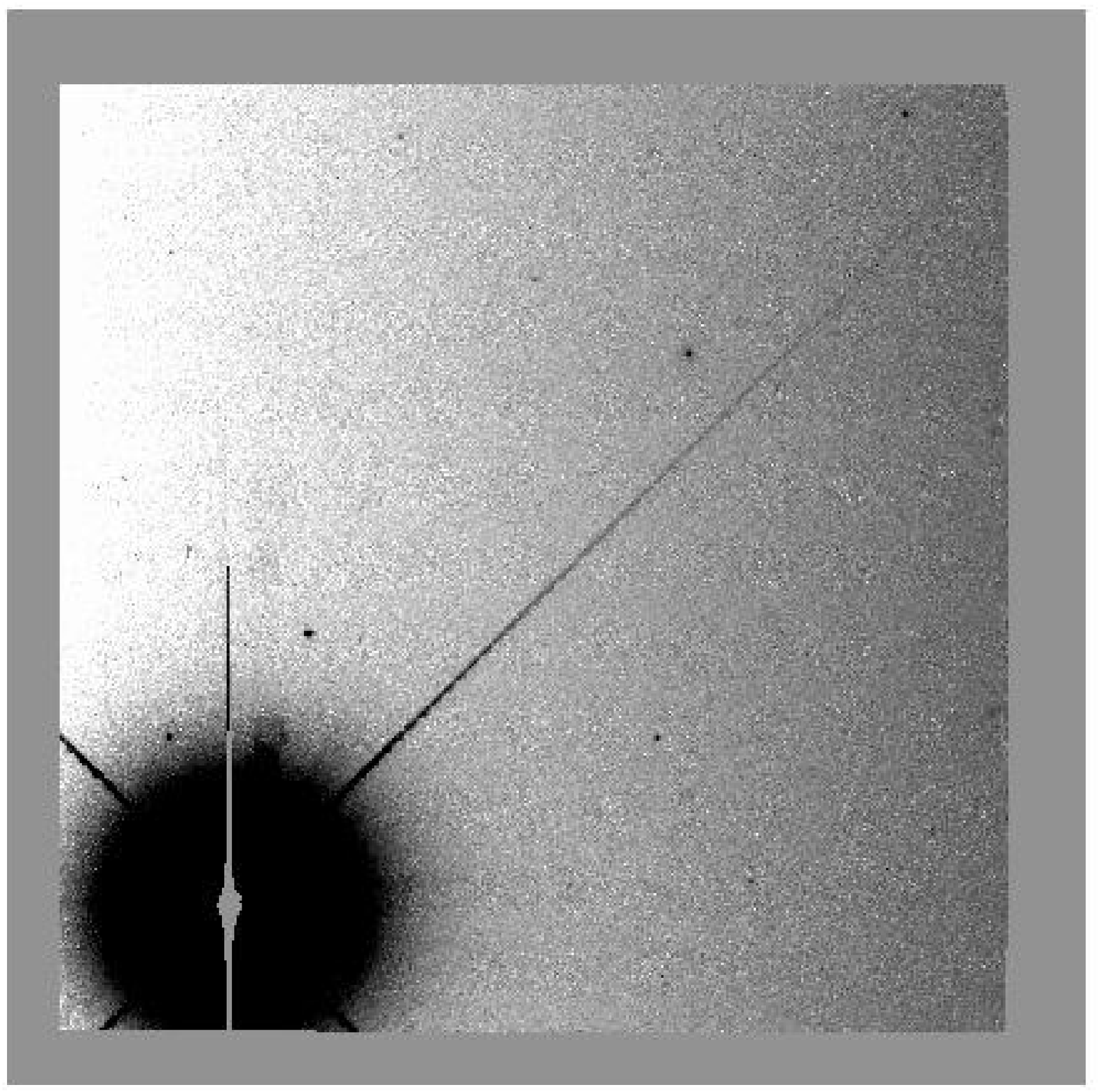} &
  \includegraphics[width=4cm]{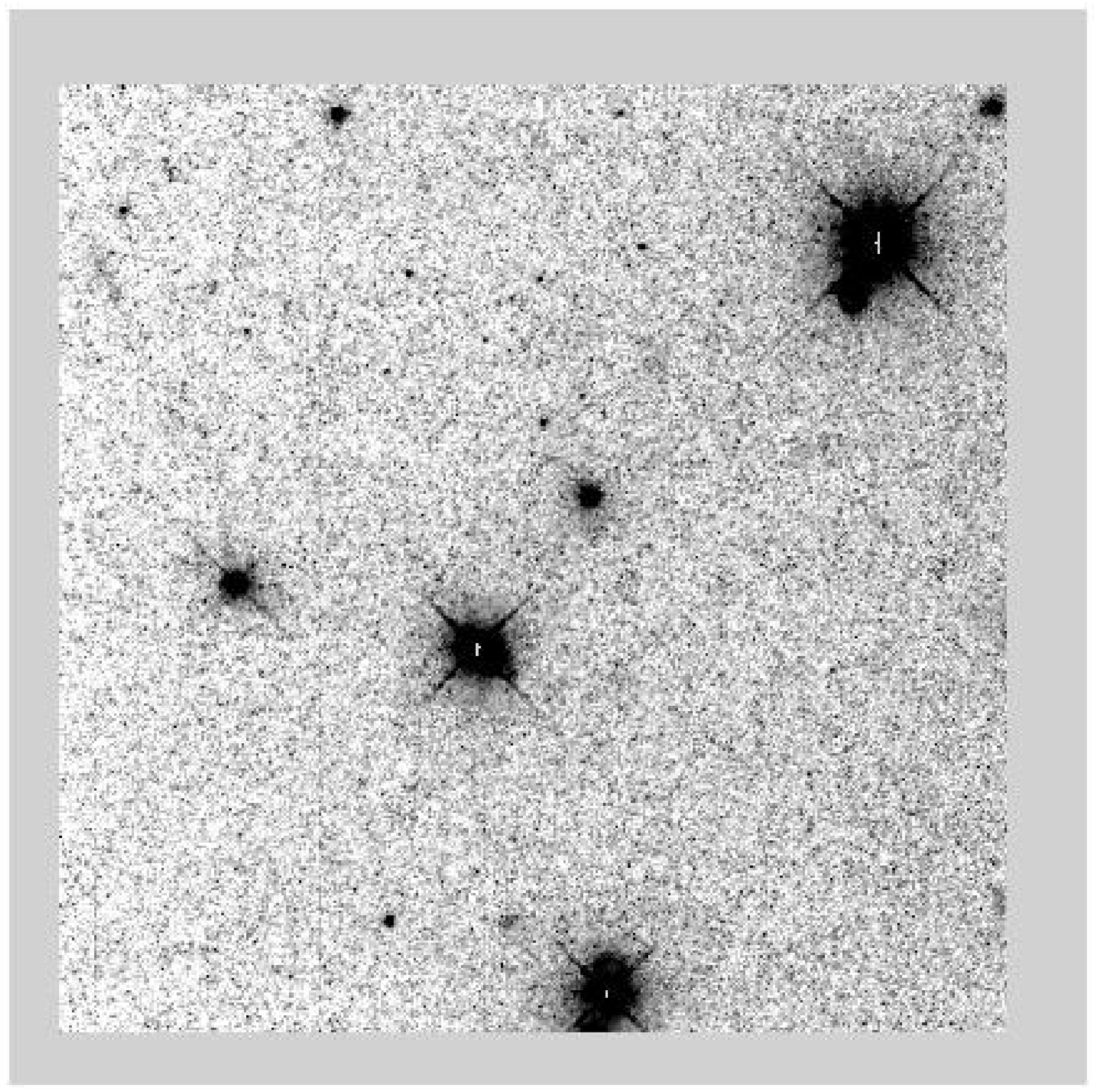} \\
  \includegraphics[width=4cm]{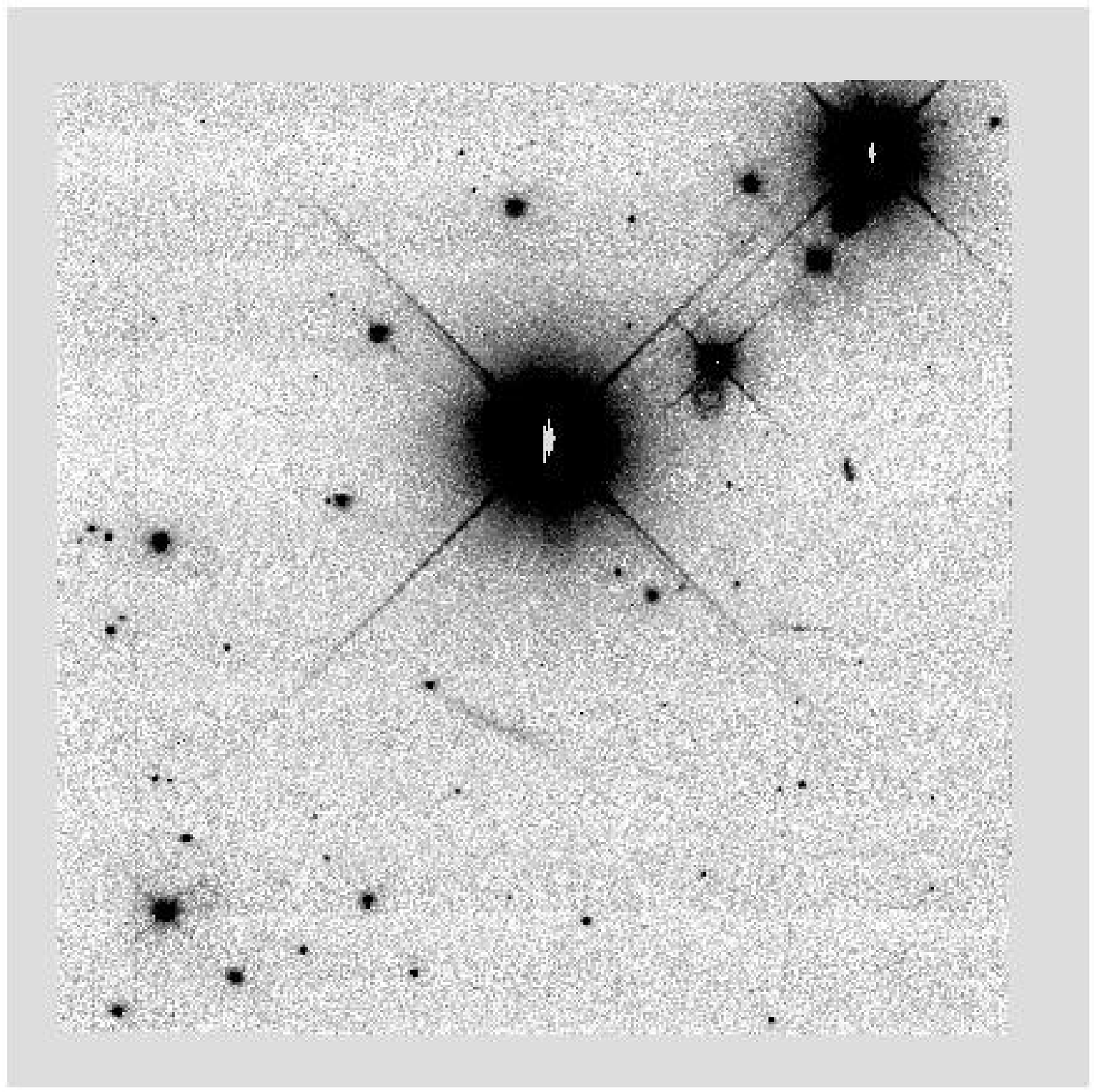} &
  \includegraphics[width=4cm]{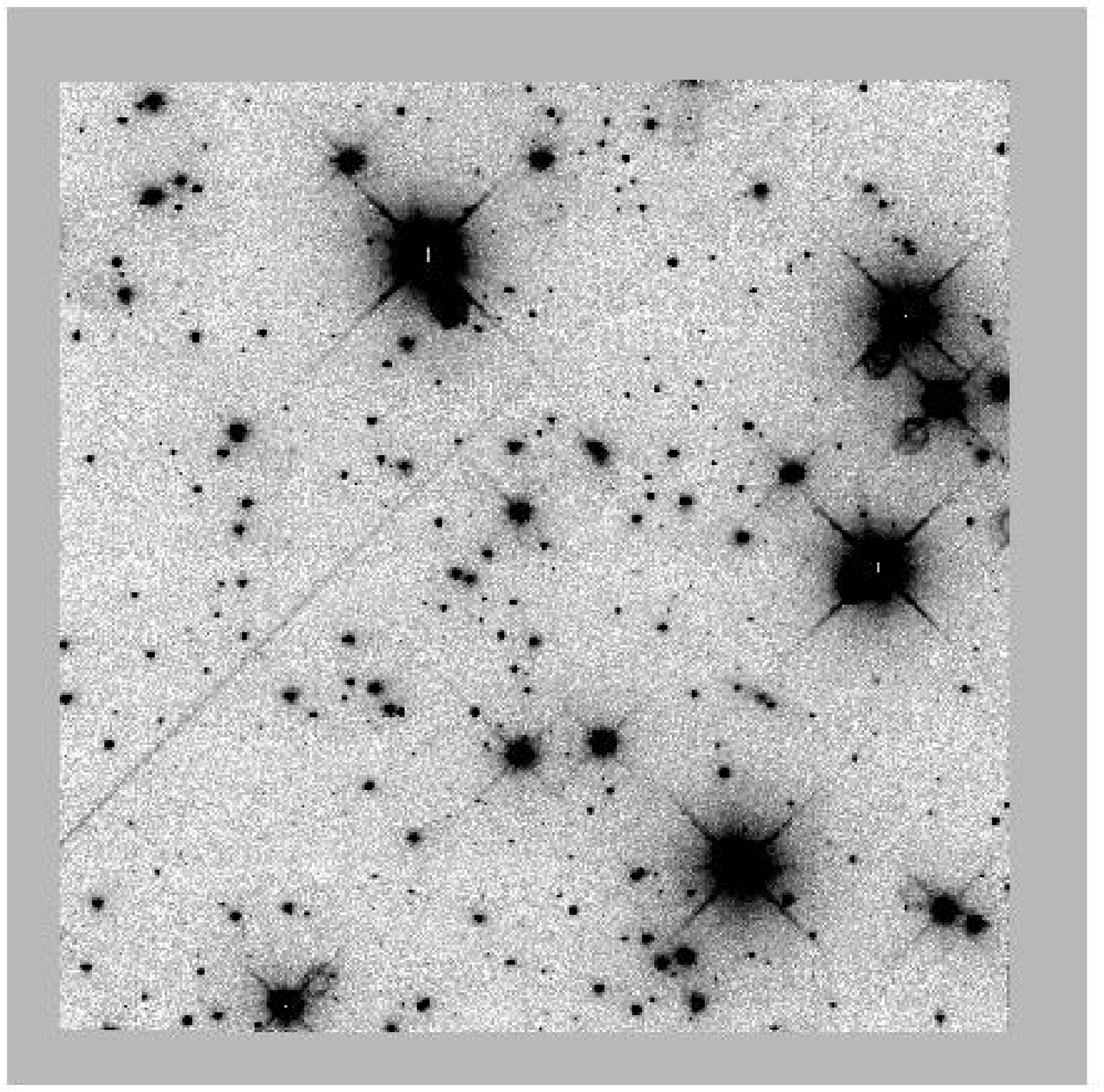} &
  \includegraphics[width=4cm]{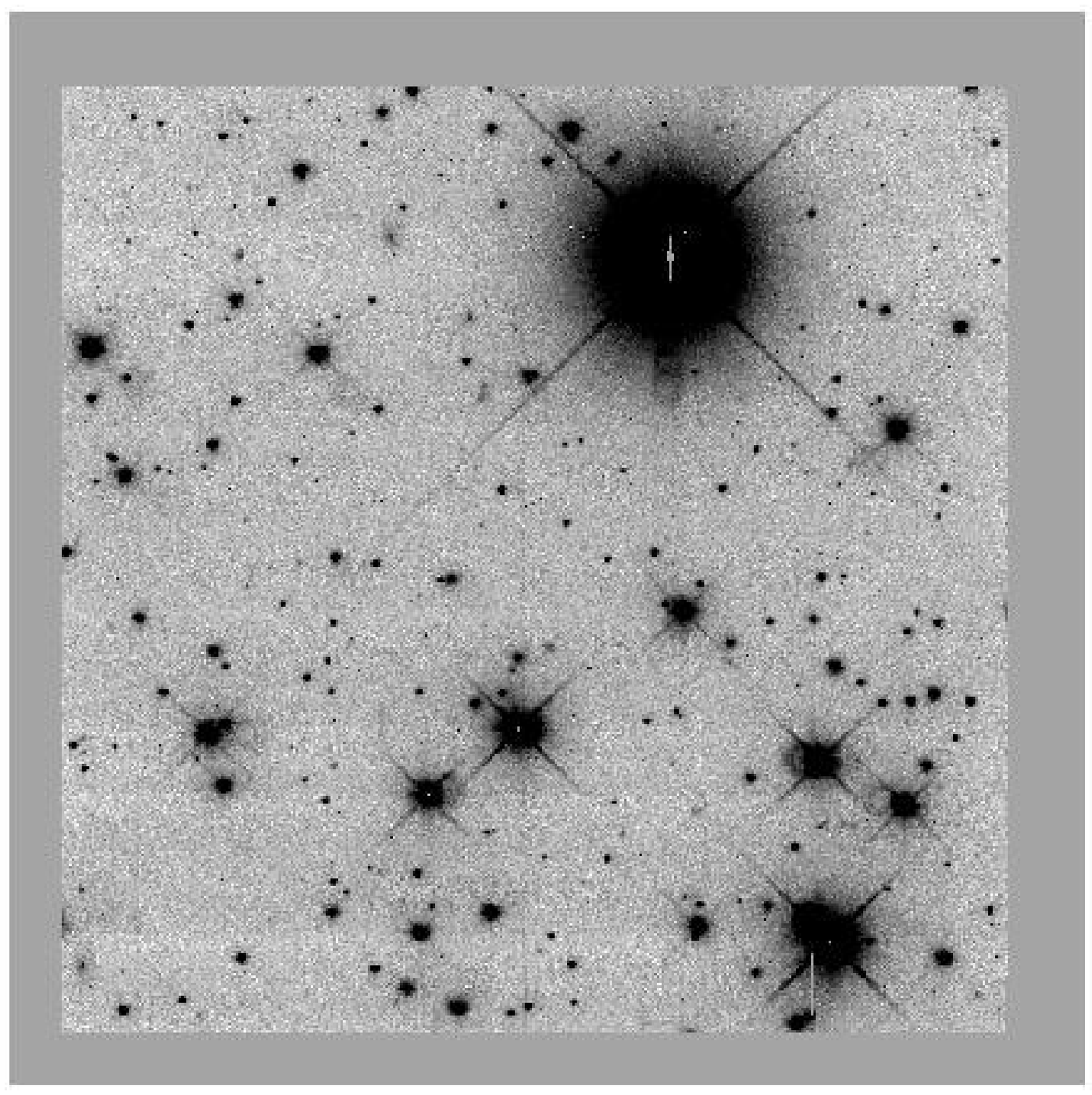} \\
  \includegraphics[width=4cm]{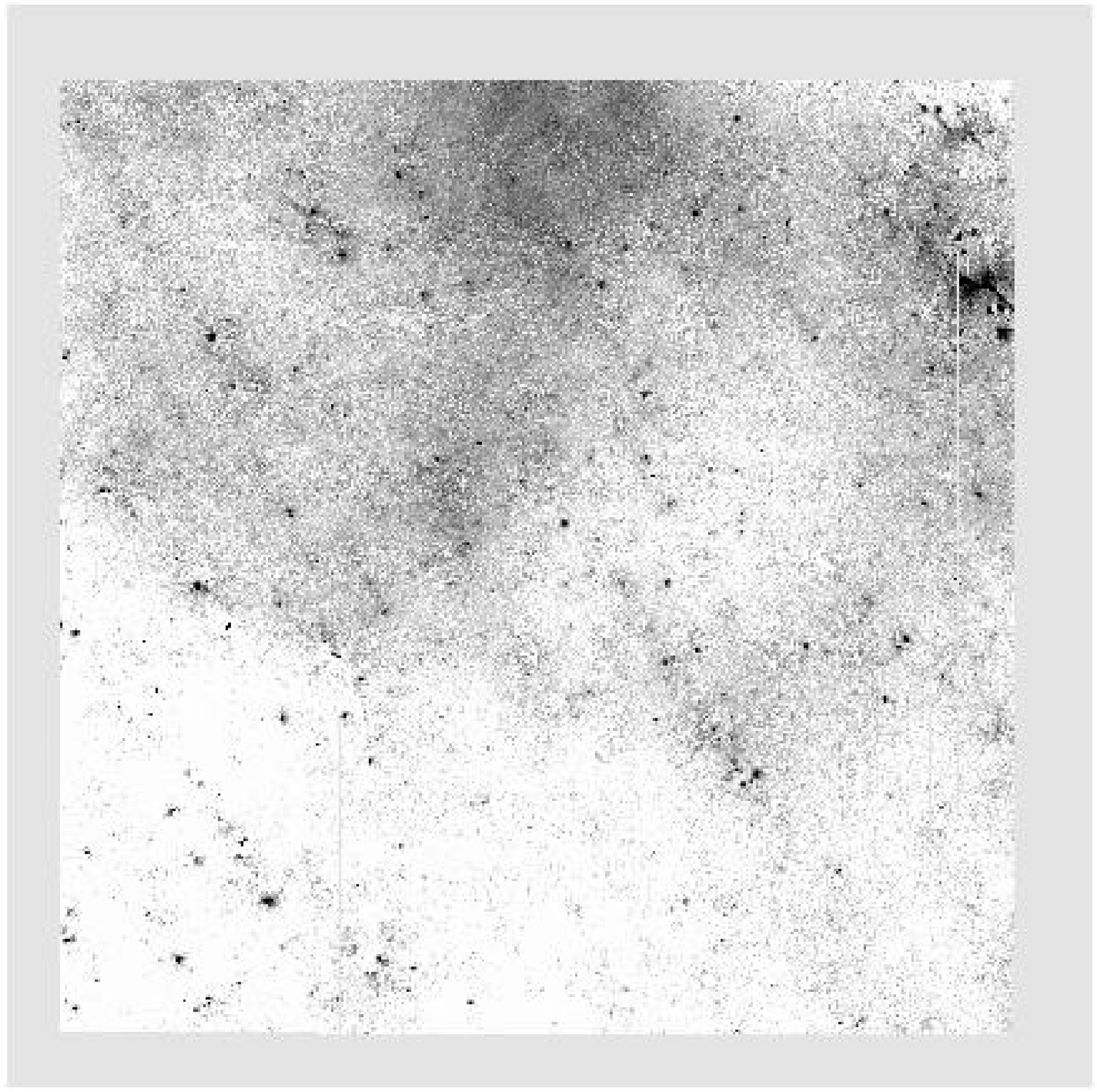} &
  \includegraphics[width=4cm]{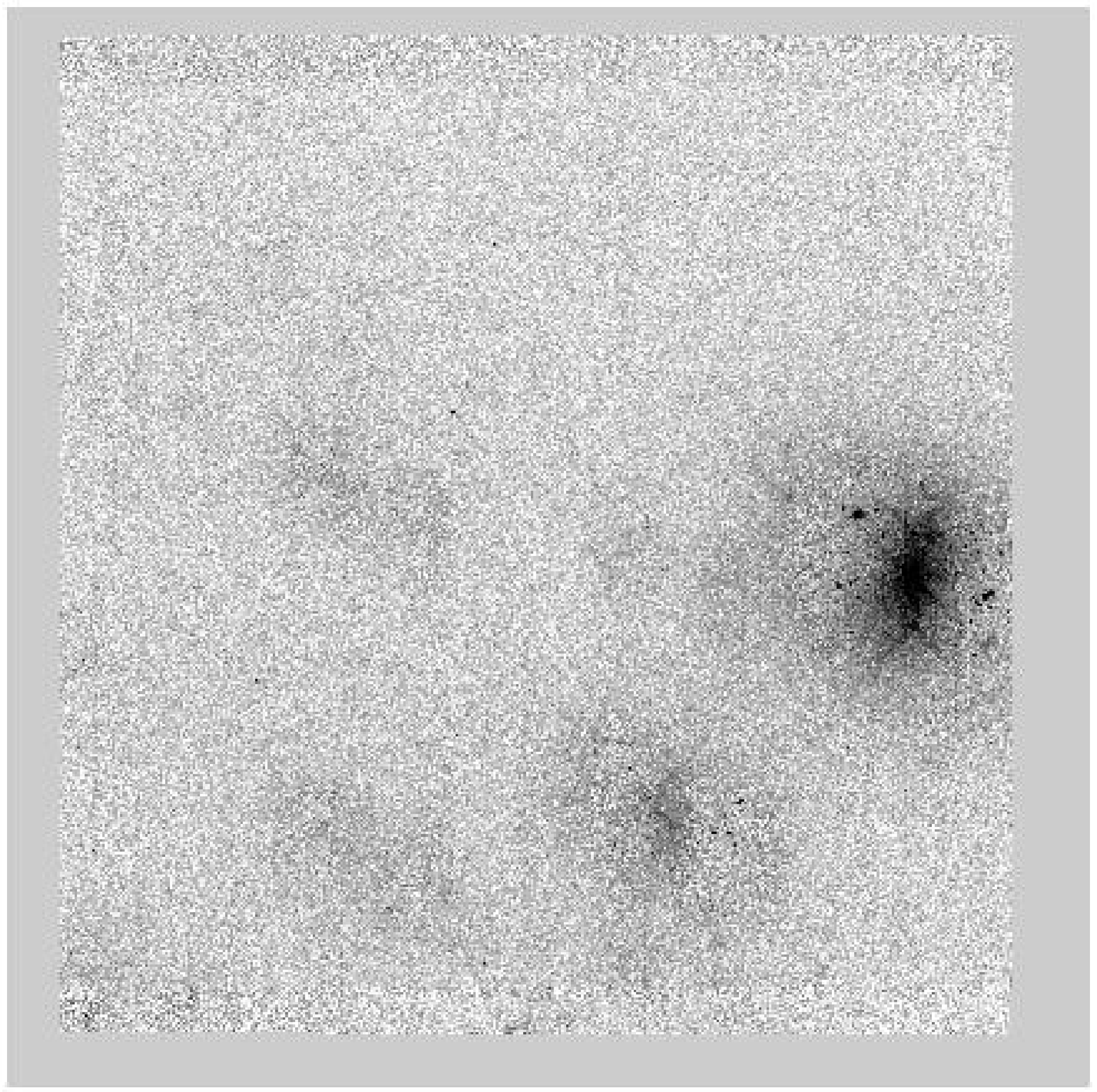} &
  \includegraphics[width=4cm]{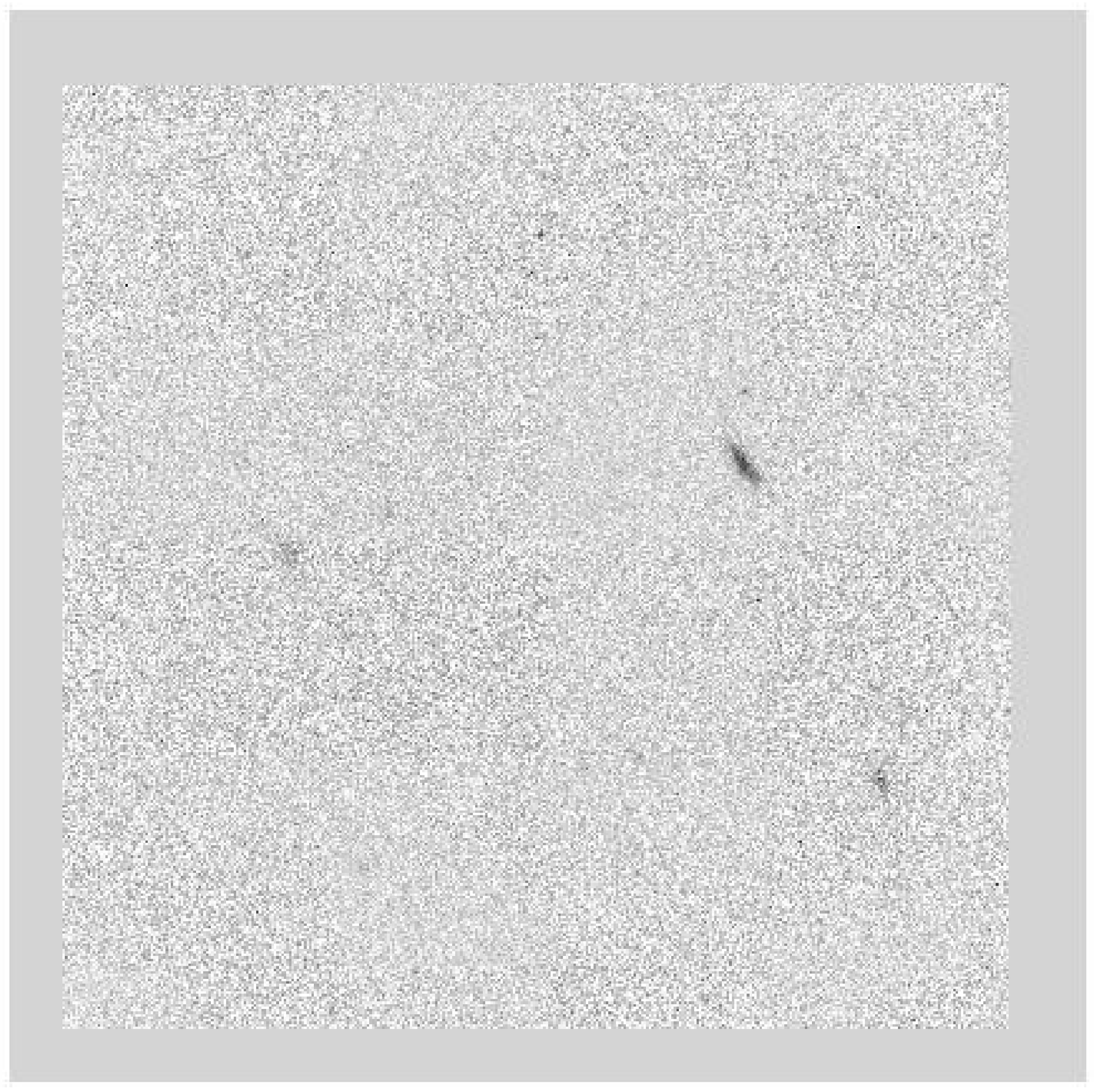} \\
  \end{tabular}
 \caption{Some examples of fields which were rejected after a visual
 inspection: the fields in the top row are dominated by one or a few
 big objects and are otherwise almost empty. In the star fields in the
 middle row there are many stars with diffraction spikes. The last
 row shows images with bad noise properties and one empty field. }
 \label{fig:notused}
\end{figure*}                               

Furthermore we did not use more than one association (data taken in a
single visit) at a particular pointing, selecting the one with the
highest S/N.   We only
coadded images taken during the same visit for which  the same guide
stars were used so that we could verify that no rotation exists
between successive images. These fields were used, however, to test
the consistency of the shape measurement, see  Sect.~\ref{ss:smearing}.

The fields were analysed using both SExtractor (\cite{BA96}~1996) and  a
modified version of the IMCAT package (\cite{KSB};
\cite{E01}~2001) using the following procedure:
\begin{enumerate}
\item SExtractor catalogues were produced using the parameter file
described in Sect.~5.2 of \cite{PaperI} (see also:
http://www.stecf.org/projects/shear/). In order to select
well-measurable objects in the  catalogues produced by SExtractor, we
applied a `flag selection', disregarding all objects which were
flagged internally by SExtractor due to problematic deblending and/or
thresholding. Furthermore, we did a visual inspection of all the
fields, masking suspect sources like galaxies with clearly separated
sub-components and 
bright stars with diffraction spikes and ghosts.
\item IMCAT catalogues were produced simultaneously. Since IMCAT was
developed specifically  for shape measurements but not for a
reliable detection of sources, we  performed a `two-code
selection': IMCAT and SExtractor catalogues were merged 
requiring an object to be detected uniquely by both codes,
which means that no object had more than one counterpart in  any of
the two catalogues in a radius  of 125~mas (5 subsampled STIS pixels)
for the central coordinates. The number of objects in the merged
catalogue diminishes by no more than 5 objects with respect to the
SExtractor catalogue for the galaxy fields and includes 98\% of objects
for the star fields. The objects which were rejected were either
flagged by IMCAT (near the border, negative flux or half-light radii)
or there were multiple detections by IMCAT where there was only one in
SExtractor, which is typical for galaxies with resolved star forming
regions.
\item In the final merged catalogue we used size and shape parameters
from IMCAT, since it was designed specifically to measure  robust
ellipticities for faint galaxy images and allows for the correction of
measured image ellipticities for shape distortion introduced by the
PSF (\cite{LK97}~1997; \cite{Hea98}~1998).  The position and
magnitude were estimated with SExtractor where we used
MAG\_ZP=26.38 which is the AB magnitude zeropoint for the STIS CLEAR
filter mode. The objects retained typically have a S/N of more than
5. The signal-to-noise ratio is measured with the \texttt{snratio}
parameter of IMCAT (see \cite{E01}~2001). 
\end{enumerate}

After catalogue production, we rejected fields with fewer than 10
(100) objects for galaxy (star) fields, which leaves us with 121 and
51 fields, respectively.

To select stars, we used size vs. S/N plots in which the stars
populate a well-defined strip, as can be seen for two examples of
star fields in Fig.~\ref{fig:rhmag}. Since the images have very
different exposure  times, we decided to use S/N rather than magnitude
to be able to use common criteria for all fields.  For the rest of the
paper we assume that stars have $2.1<\rh <2.6$ in subsampled STIS
pixels ($52 \mathrm{mas} <\rh < 65 \mathrm{mas}$) and
$\mathrm{S/N}>10$; galaxies are selected with $\rh >2.6$. The
half-light radius is measured with IMCAT and is the radius at which
half of the flux of an object is included. 

The mean number of selected galaxies per association is 18 on
the galaxy fields (25/arcmin$^2$), which is similar to the galaxy number
density found by \cite{Rea01}~(2001) in the `Groth Strip' with
$\mathrm{I}<26\mathrm{mag}$;  however, the way galaxies are selected in
their paper is different from ours.

\begin{figure}                             
 \resizebox{\hsize}{!}{\includegraphics{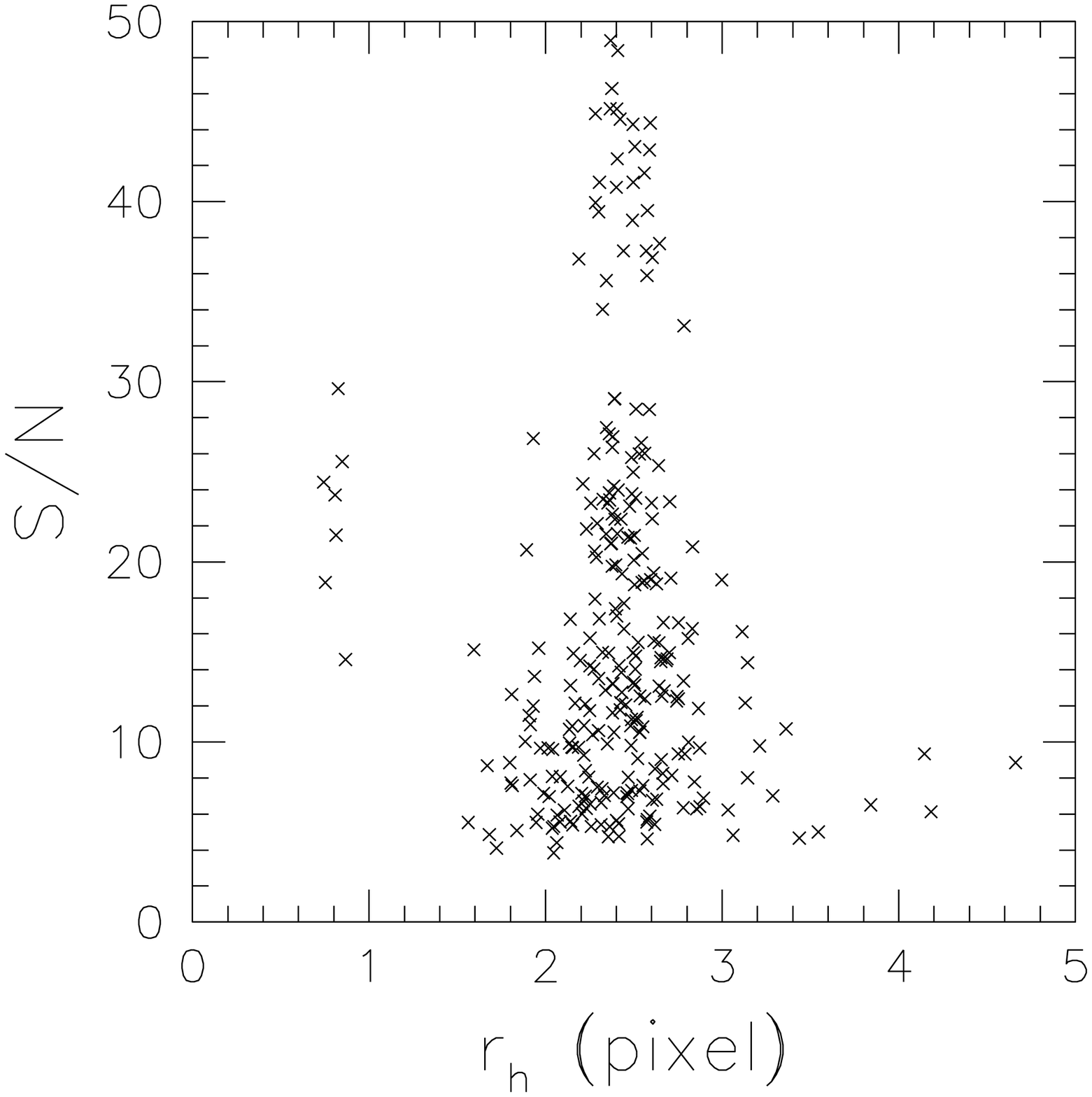}
		       \includegraphics{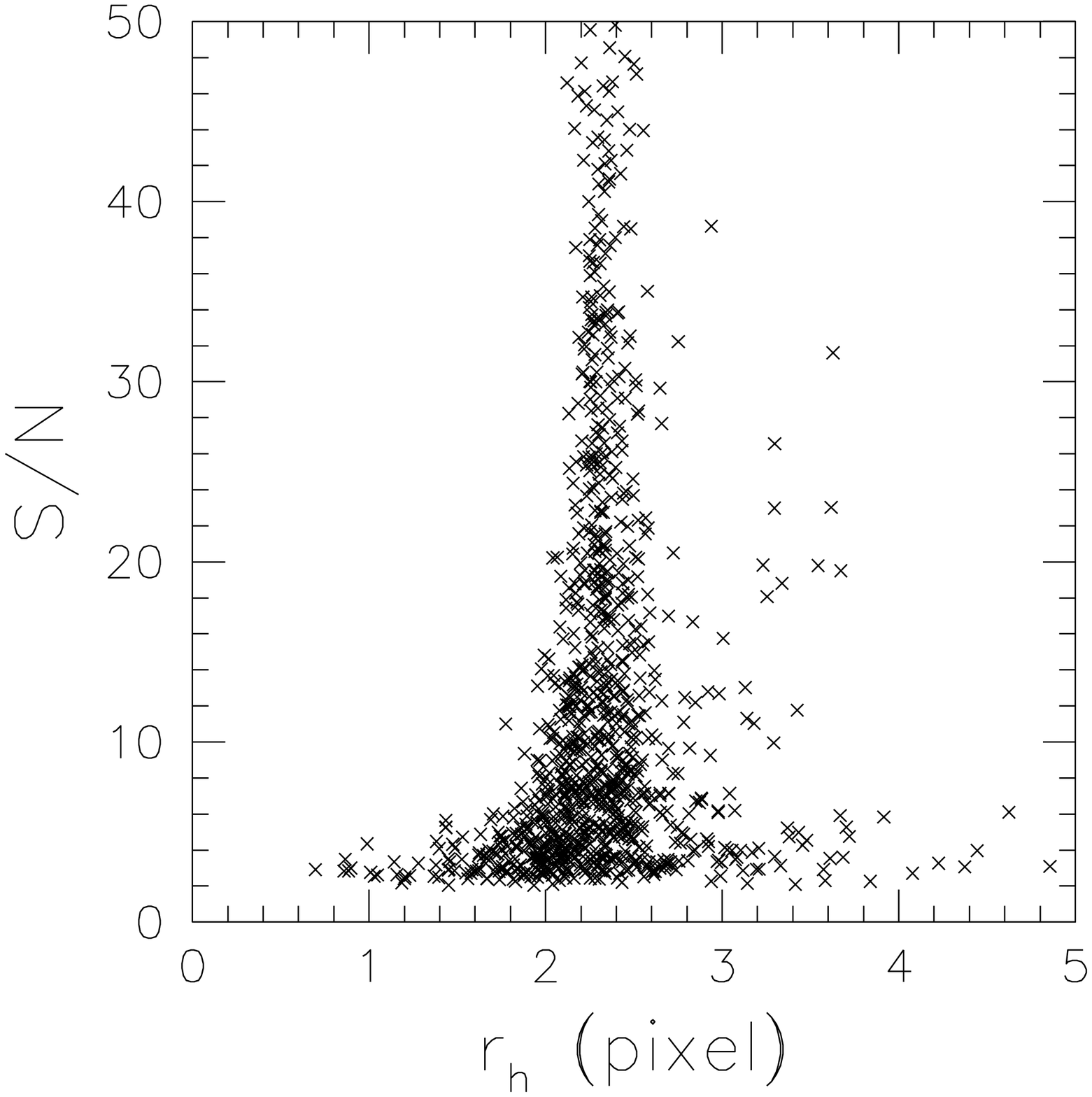}}
 \caption{Size vs.\ S/N diagrams for the star fields o3zf01010\_3\_ass
 (left) and  o48b41010\_3\_ass (right).  The half-light radius $\rh$ is  
 given in subsampled STIS pixels. One can clearly distinguish the strip 
 populated by stars around $\rh=2.3$ (57.5mas). The strip at
 $\approx$0.8 pixels  seen in the left panel is due to single noisy 
 pixels (see also Fig.~6 of \cite{PaperI}).}
 \label{fig:rhmag}
\end{figure}                               

\end{section}

\begin{section}{Analysis of PSF anisotropy}\label{sc:PSF}
Our main scientific motivation for using data from the STIS Parallel
Survey is their suitability for the detection and measurement of
cosmic shear. Since the  
expected distortion of image ellipticities on the STIS angular scale
is a few percent, any instrumental distortion and other causes of PSF
anisotropy need to be understood and controlled to an accuracy of
$\lesssim 1\%$. In particular, the PSF anisotropy needs to
be either small in amplitude or stable in time. 

The shape of the PSF anisotropy is estimated using the \cite{KSB}
ellipticity parameters. These describe the elongation and orientation
of the equivalent ellipse which best reproduces the PSF shape, as
determined from the weighted second order brightness moments,
\begin{equation}\label{eq:Qij}
 Q_{ij}=\int \mathrm{d}^2\theta \, W(\theta) \theta_i \theta_j
 f(\vec{\theta}),
\end{equation}
where $\vec{\theta} = (\theta_1,\theta_2)$, $\theta = |\vec{\theta}|$,
and angles are measured relative to the centroid
$\vec{\theta_\mathrm{c}}$ of the object, as defined by 
\begin{equation}
 \int \mathrm{d}^2\theta \, W(\theta_\mathrm{c})\theta_i 
 f(\vec{\theta_\mathrm{c}}) = 0.
\end{equation}
$f(\vec{\theta})$ is the observed surface brightness and $W(\theta)$
is a weight function, which we take to be a Gaussian with an
appropriate filter scale. This scale is obtained by determining the
significance of detecting a peak for a range of filter scales and
using the scale which gives maximum significance as described in
\cite{KSB}.  

The complex ellipticities are then defined as 
\begin{equation}
 \e=\frac{Q_{11}-Q_{22} + 2\mathrm{i}Q_{12}} {Q_{11}+Q_{22} } ;
\end{equation}
for an object with elliptical isophotes
\begin{equation}
 \e =  \frac{1-r^2}{1+r^2} \, \mathrm{e}^{2 \mathrm{i}\vartheta}
\end{equation}
where $r=b/a$ is the axis ratio of the corresponding ellipse and its
position angle is $\vartheta=0.5\arctan{\e_2/\e_1}$.

In Fig.~\ref{fig:aniexp} the mean values of the two ellipticity
components of stars over the whole field are shown as a  function of
the exposure date. The mean ellipticity of all fields is 1.5\% for
$\e_1$ and 0.5\% for $\e_2$ with a dispersion of 1\%, which is
sufficiently small for our cosmic shear analysis, as will be shown in
Sects.~\ref{ss:cPSF} and \ref{ss:rescorr}. Only a small fraction of
the scatter is due to the different distribution of stars over the
fields. If we divide the
star fields into time intervals, the mean ellipticities agree with each
other on the $1\sigma$ level, therefore the anisotropy can be
considered to be constant over the time period covered. 

\begin{figure}
 \resizebox{\hsize}{!}{\includegraphics{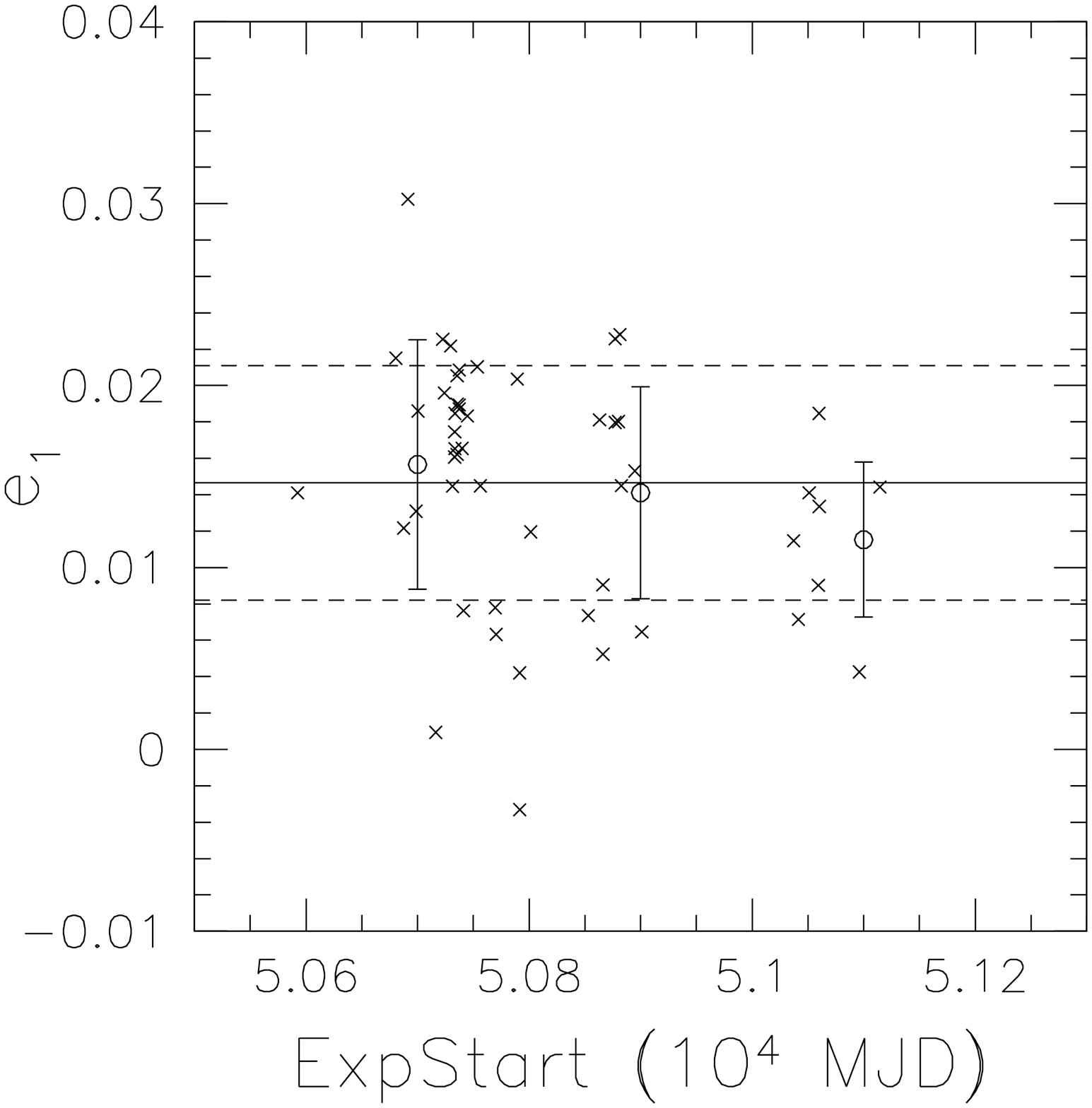}
	               \includegraphics{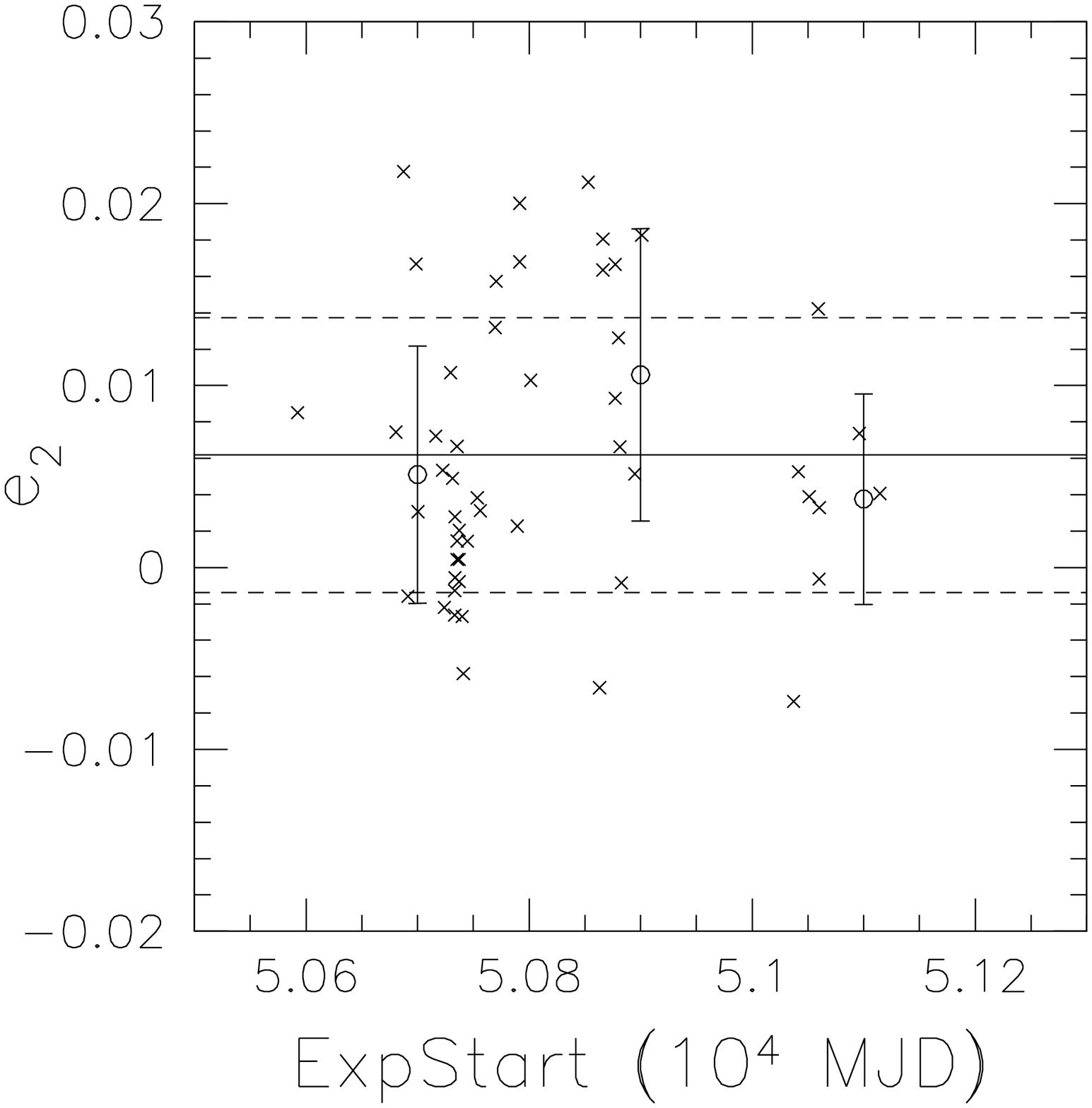}}
 \caption{Mean of the ellipticity components $\e_1$ (left) and $\e_2$
 (right) of the star fields  vs.\ exposure start time (Modified 
 Julian Date).  The  straight lines show the mean over all the fields, 
 the dashed lines show the $1\sigma$ dispersion. The circles show 
 the mean over stars in bins between $5.06\times 10^4$, 
 $5.08\times 10^4$, $5.10\times 10^4$, $5.12\times 10^4$ (MJD) with 
 the error bars showing the 1$\sigma$ dispersion.  }
 \label{fig:aniexp}
\end{figure}                               

In addition to the variation from field to field (i.e., in time), we
also find a spatial  variation of the PSF within individual
fields. This effect is shown for two fields in Figs.~\ref{fig:PSFani1}
and \ref{fig:PSFani2}, bottom left panels. We fit the ellipticities
with a second order polynomial at the position $\vec{\theta}=(x,y)$ on
the CCD:
\begin{equation}
 \e_\alpha(\vec{\theta}) = a_{\alpha0} + a_{\alpha1}x + a_{\alpha2}x^2
           + a_{\alpha3}y + a_{\alpha4}y^2 + a_{\alpha5}xy,
\end{equation}
where $\alpha=1,2$.  For the fitting an iterative procedure was used:
for stars selected by $\rh$ and S/N, as described in
Sect.~\ref{sc:cat}, a first guess of the polynomial was calculated
by solving the normal equations.  From this initial fit, a dispersion
was calculated by
\begin{equation}
 \sigma^2_\alpha = \frac{\sum_i (\e_\alpha(\theta_i) - \e_{\alpha
 i})^2} {N_\mathrm{star}-N_\mathrm{coeff}} .
\end{equation}
Only stars for which the ellipticity components did not deviate more
than $3\sigma$ in both $\e_1$ and $\e_2$ from the first fit were used
for the second fit. The $3\sigma$ clipping was then applied to all the 
preselected stars (not only the left-overs from the first fit) to
avoid biasing. This  procedure was repeated until it converged, 
which was typically after no more than 10 iterations, with 5--10\%
outliers disregarded.  For the two star fields in Fig.~\ref{fig:rhmag},
the ellipticities and the values for the fitted polynomials at the
star positions are shown in Figs.~\ref{fig:PSFani1} and \ref{fig:PSFani2}.

For the anisotropy correction of the galaxies each star field
polynomial is applied to each galaxy field [see Sect.~\ref{ss:cPSF}]. 
The time period  when the galaxy fields were observed is well covered
by the observation times of the star fields. 

\begin{figure}                             
 \resizebox{\hsize}{!}{\includegraphics{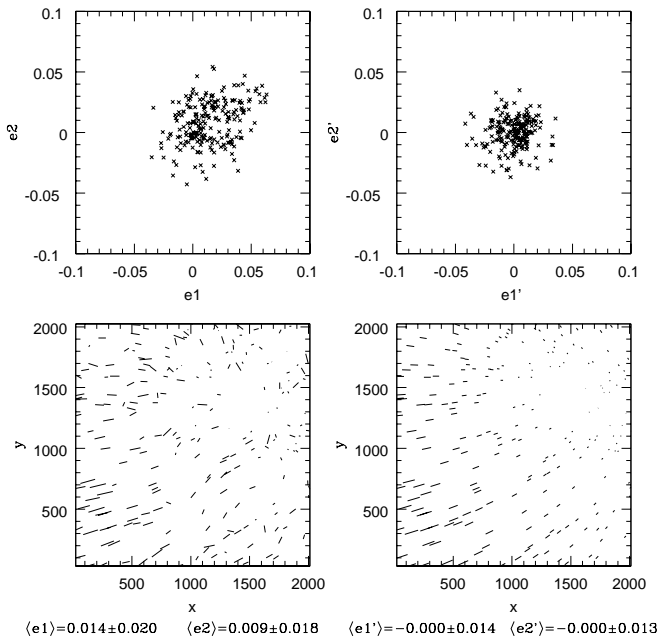}}
 \caption{For the star field o3zf01010\_3\_ass the distribution of the
 ellipticities of  stars are shown before (top left) and after (top 
 right) correcting for PSF anisotropy. The bottom left panel shows 
 the spatial distribution of the ellipticities across the STIS field, 
 the bottom right the fitted second-order polynomial at the star  
 positions. The length of the sticks indicates the modulus of the 
  ellipticity, the orientation gives the position angle. The mean 
 ellipticities before and after the  correction are given at the 
 bottom. Note that the mean ellipticity after the correction is zero 
 and that the dispersion  decreases. }   
 \label{fig:PSFani1}                              
\end{figure} 
                              
\begin{figure}                             
 \resizebox{\hsize}{!}{\includegraphics{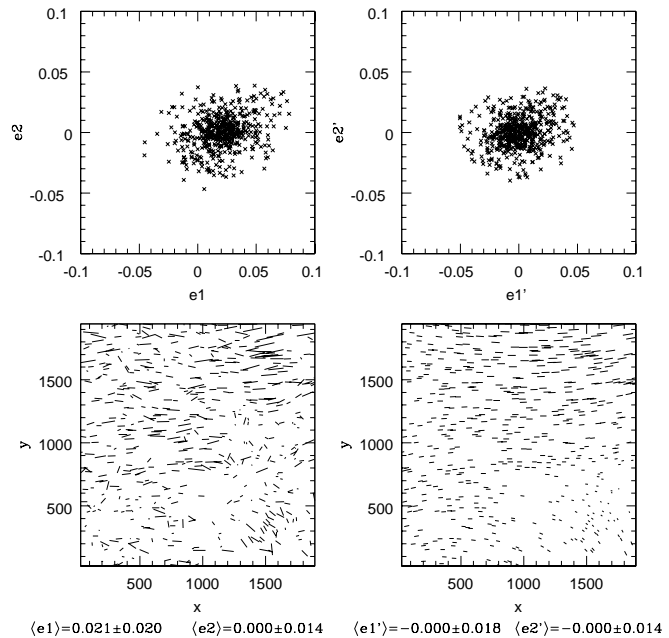}} 
 \caption{Same as Fig.~\ref{fig:PSFani1} for the star field 
 o48b41010\_3\_ass.}
 \label{fig:PSFani2}                              
\end{figure}                               

To check the stability of our coaddition procedure, we also analysed
individual exposures with respect to PSF size and anisotropy. The
stars in the individual exposures were selected by $1.1<\rh<1.4$ STIS
pixels and $\mathrm{S/N}>10$. We find  that the stars in the drizzled images
are slightly larger ($\approx$0.1 subsampled STIS
pixels) than in the individual exposures [see Table~\ref{tab:PSFsize}]. 
The broadening of the PSF is a known property of drizzle if one uses
PIXFRAC=1 (see \cite{FH98} 1998). 
In Sect.~\ref{ss:sim} we describe some simulations
carried out to check our analysis of the cosmic shear; in the
simulated star fields we find the same increase in sizes of stars
($\approx 5\%$) from the individual images to the drizzled ones, as we
observed in the archive data. 

From the results in Table~\ref{tab:PSFsize} we find more stars in the
drizzled images than in the undrizzled ones. For associations where
many individual exposures have been coadded this can be naturally
attributed to a gain in S/N. However, we also find more objects if just
one individual image is drizzled, as can be seen in the last two rows
of Table~\ref{tab:PSFsize} for the star field o4xcll010\_1\_ass. This
is a result of the  combination of drizzling and IMCAT: an object
which has a size of only one STIS pixel or slightly larger is distributed
over more  subsampled STIS pixels in the drizzling process. Therefore an
object which was rejected by IMCAT on the undrizzled image because of
its size of only $\approx 1$ pixel (and therefore high probability to
be noise) is now detected significantly.

\begin{table}
\center
\caption{Mean half-light radius and error on the mean for individual
exposures (in STIS pixels) and for the \textit{corresponding coadded
association  (in subsampled STIS pixels)} for all stars on an
image. The first Col. identifies the individual exposure or the
coadded image. The second Col. yields the number of stars used to
calculate the mean. The third and fourth Cols. show the mean
half-light radius and the error on the mean for individual exposures
in STIS pixels (50 mas) and for  the coadded images in subsampled STIS
pixels (25 mas).  The last two Cols. show the date of observation
$t_\mathrm{date}$ (MJD) and the exposure time $t_\mathrm{exp}$ in
seconds.  }
\label{tab:PSFsize}
\begin{tabular}{|l|rrrrr|}
\hline 
image & N & $\ave{r_\mathrm{h}}$ & $\sigma$ & $t_\mathrm{date}$ & $t_\mathrm{exp}$ \\
      &   & image    &               & (MJD)    & \\
      &   & pixels   & $\times 10^3$ & 50000$+$ & sec \\
\hline  
 o3zf01010         & 132 & 1.1503 & 6.8 & 592.4319  & 60  \\
 o3zf01020         & 139 & 1.1563 & 6.9 & 592.4344  & 60  \\
 o3zf01030         & 135 & 1.1389 & 6.4 & 592.4369  & 60  \\
 o3zf01040         & 132 & 1.1296 & 5.9 & 592.4415  & 60  \\
 o3zf01090         & 148 & 1.1021 & 5.2 & 592.4592  & 64  \\
 o3zf010a0         & 158 & 1.0989 & 5.2 & 592.4634  & 72  \\
 o3zf010b0         & 151 & 1.1505 & 6.0 & 592.5012  & 72  \\
 o3zf010c0         & 158 & 1.1357 & 6.6 & 592.5039  & 72  \\
 o3zf010d0         & 147 & 1.1345 & 5.2 & 592.5087  & 72  \\
 o3zf010e0         & 155 & 1.1219 & 5.8 & 592.5114  & 72  \\
 o3zf010f0         & 151 & 1.1227 & 5.7 & 592.5166  & 72  \\
 o3zf010g0         & 147 & 1.1040 & 5.0 & 592.5192  & 72  \\[1.5ex]
 \textit{o3zf01010\_3} & \textit{322} & \textit{2.4016} & \textit{6.3} & \textit{592.4319}  & \textit{808 }\\ 
\hline				   	   
 o48b3w010         & 435 & 1.1142 & 3.9 & 735.3806  & 400 \\
 o48b51010         & 430 & 1.1237 & 4.1 & 737.1772  & 400 \\
 o48b5d010         & 420 & 1.1129 & 3.8 & 737.3673  & 400 \\
 o48b5p010         & 397 & 1.1044 & 4.2 & 737.6852  & 400 \\[1.5ex]
 \textit{o48b3w010\_3} & \textit{660} & \textit{2.3397} & \textit{4.6} & \textit{735.3806}  & \textit{1600}\\ 
\hline				   	   
 o4xcll010         & 432 & 1.1631 & 3.3 & 1096.1600  & 300 \\[1.5ex]
 \textit{o4xcll010\_1} & \textit{467} & \textit{2.4238} & \textit{5.8} & \textit{1096.1600}  & \textit{300 }\\
\hline
\end{tabular}
\end{table}

For the individual undrizzled images of some of the star fields, we
also analysed the PSF anisotropy and we find very short timescale
variations of the anisotropy pattern over the fields, as can be seen
in Fig.~\ref{fig:indiv} for two individual exposures of the star field
o3zf01010\_3\_ass.  The short timescale variations of the PSF
anisotropy pattern are likely to be due to ``breathing'' of the
telescope. However, they do not affect our analysis, as will be shown
in Sect.~\ref{ss:cPSF}.

\begin{figure}
 \resizebox{\hsize}{!}{\includegraphics{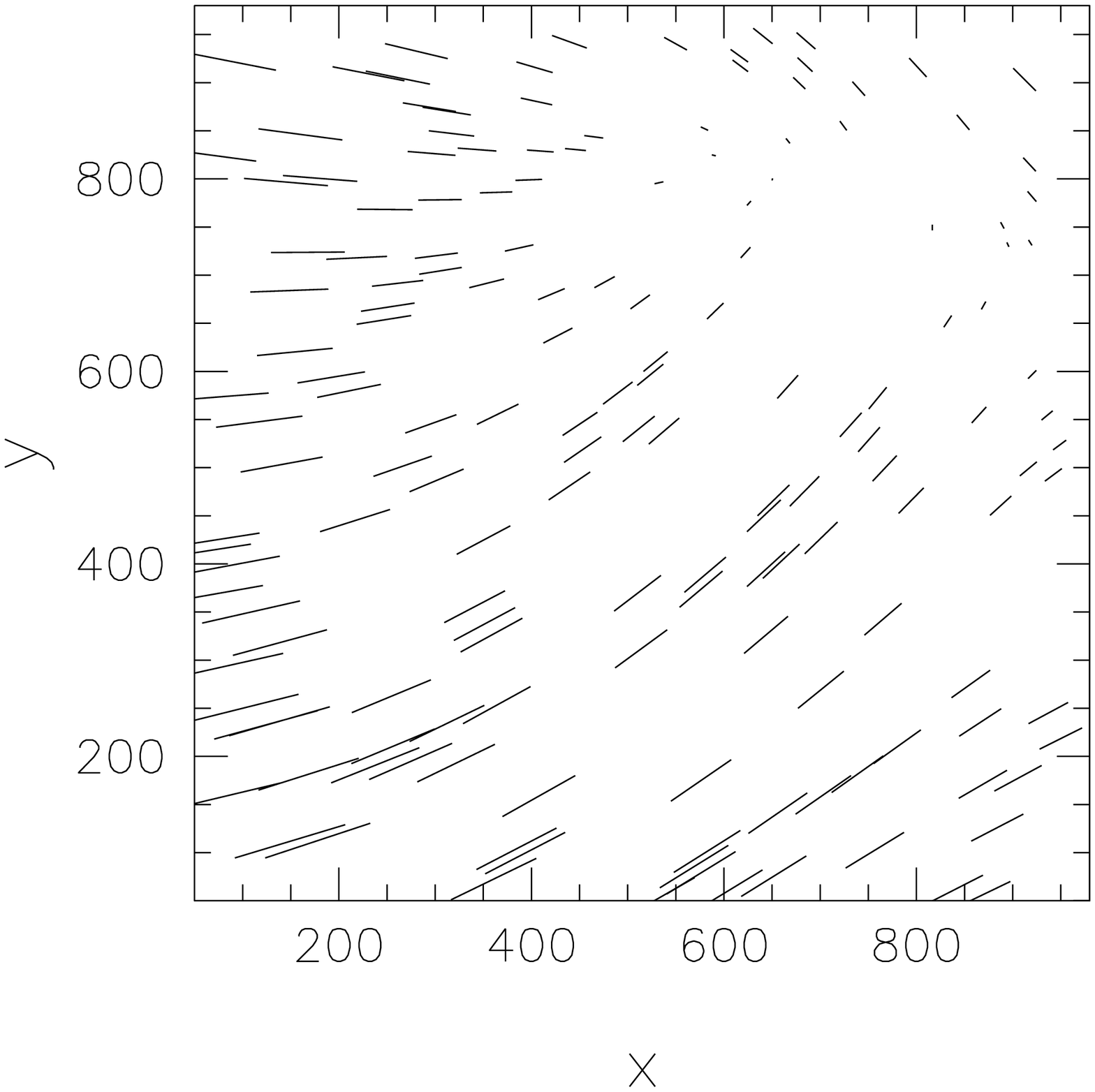}
		       \includegraphics{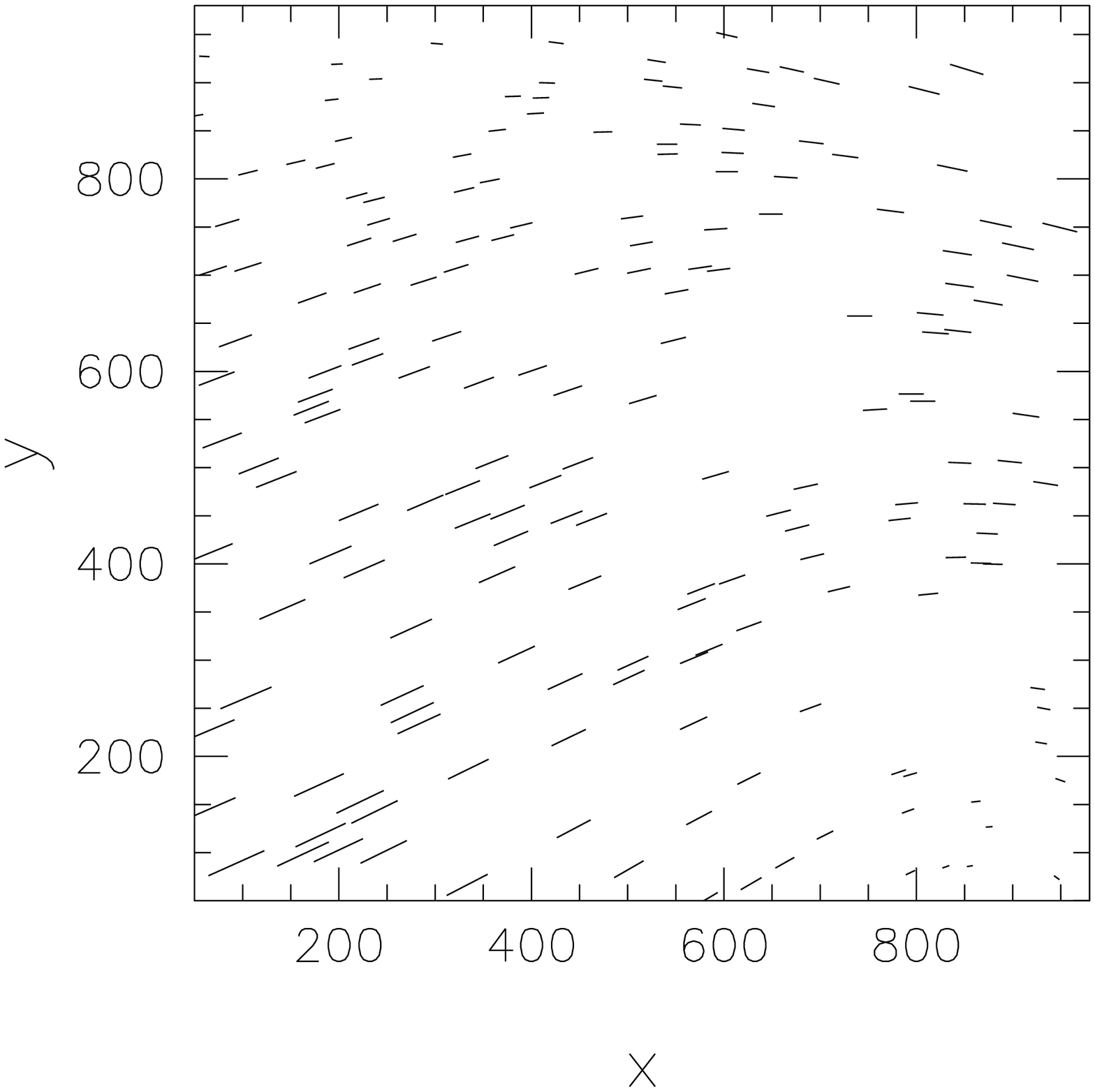}}
 \caption{We show the values of the polynomials for the anisotropy
 correction of the two individual (undrizzled) exposures o3zf01040 
 (left) and o3zf01090 
 (right) which are members of the association o3zf01010\_3\_ass shown 
 in Fig.~\ref{fig:PSFani1}. The two individual fields were  taken with 
 a time difference of only 30 minutes and demonstrate the very short 
 timescale variations of the PSF anisotropy pattern.  }
 \label{fig:indiv}
\end{figure}

\end{section}

\begin{section}{Shear analysis}\label{sc:shear}

\begin{subsection}{Cosmic shear theory}\label{ss:theory}
The rms shear in a circular aperture with angular radius $\theta$ is
related to the power spectrum  of the surface mass density $\kappa$ by
\begin{equation}\label{eq:cstheory}
 \cs = 2\pi \int_0^\infty \mathrm{d}s \, s \, P_\kappa(s) \left[
 I(s\theta) \right]^2,
\end{equation}
where $I(\eta):=\mathrm{J}_1(\eta)/(\pi \eta)$ and $\mathrm{J}_1$ is
the Bessel function of  the first kind. $P_\kappa$ is in turn related
to the three-dimensional power spectrum by a simple projection (see
e.g.\ \cite{Bea91}~1991, \cite{K92}~1992, \cite{Schneider98a}~1998a,
\cite{M99}~1999, \cite{BS01}~2001). A similar formula for a square
aperture is given by \cite{K92}~(1992). 

Eq.~(\ref{eq:cstheory}) is correct if one measures the shear in a
circular aperture. To quickly compare our result from the square
STIS field to theoretical predictions, we calculate the theoretically
expected value for a circular field with the  same area, which would
then have an effective radius of about 30\arcsec. The error from this
approximation is much smaller than the statistical error bars on our
present result.

Let $\e_{in}$ denote the complex ellipticity of the $i$-th galaxy on
the $n$-th field, then the quantity we measure for each field is
\begin{equation}\label{eq:csn1}
 \csn := \frac{1}{N_n (N_n-1)} \sum_{i\neq j} \e_{in} \e_{jn}^\star
\end{equation}
where $N_n$ is the number of galaxies in the $n$-th field, or
\begin{equation}\label{eq:csn2}
 \csn := \frac{ \sum_{i\neq j} w_{in} w_{jn} \e_{in} \e_{jn}^\star}
 {\sum_{i\neq j} w_{in} w_{jn}} ,
\end{equation}
where $w_{in}$ is the weight   of the $i$-th galaxy in the $n$-th
field, and  $\e^\star$ denotes the complex conjugate. This is an unbiased
estimate of the cosmic 
shear dispersion in the $n$-th field. (Note that $\csn$ is not positive
definite.)   From this, one obtains an unbiased estimate of the cosmic
shear dispersion:
\begin{equation}\label{eq:cs1}
 \cs = \frac{1}{N_\mathrm{f}}\sum_{n=1}^{N_\mathrm{f}} \csn ,
\end{equation}
where $N_\mathrm{f}$ is the number of galaxy fields, or
\begin{equation}\label{eq:cs2}
 \cs = \frac{\sum N_n \csn}{\sum N_n} ,
\end{equation}
where we weight each field by the number of galaxies per field to
minimize Poisson noise. Since our galaxy fields have a large spread
in exposure times, and therefore in number of galaxies per field, we
use Eq.~(\ref{eq:cs2}) for our analysis.

Assuming $N_\mathrm{f}$ galaxy fields with the same number $N_n=N$ of
galaxies per field and the same redshift distribution of the sources,
the errors from the intrinsic ellipticity distribution and the cosmic
variance in the absence of kurtosis are given by 
\begin{equation}\label{eq:sigintr}
\sigma^2_\mathrm{intr} = \sqrt{2} \frac{\sigma_\mathrm{s}^2}
                                       {N \sqrt{N_\mathrm{f}}}  
\end{equation}
and
\begin{equation}\label{eq:sigcv}
\sigma^2_\mathrm{cv} = \sqrt{2} \cs /\sqrt{N_\mathrm{f}} . 
\end{equation}
With an rms intrinsic ellipticity $\sigma_\mathrm{s}\approx
\sqrt{2}\times26\%$ [see Fig.~\ref{fig:pelli}] and an expected shear of a
few percent on the STIS angular scale we find $\sigma^2_\mathrm{intr}
\approx  10^{-3}$ and $\sigma^2_\mathrm{cv} \approx 10^{-4}$, showing that
the noise from the intrinsic ellipticity distribution dominates over
the sampling variance. 

\end{subsection}

\begin{subsection}{Anisotropy correction}\label{ss:cPSF}
As shown in Sect.~\ref{sc:PSF}, the STIS PSF is remarkably symmetric,
i.e., it exhibits a very small ($\lesssim 1\%$) anisotropy component. 
Furthermore, it does not vary significantly in time ($\sigma \approx
1\%$), as shown in Fig.~\ref{fig:aniexp}. 

Of the 51 star fields, 21 were selected for having a good spatial
coverage and a small intrinsic dispersion in the ellipticities of
stars, which allowed us to obtain good fits to the anisotropy pattern,
to minimize the noise in the PSF correction.

The galaxy ellipticities were corrected for PSF anisotropy according
to \cite{KSB} as follows:
  
The total response of a galaxy ellipticity to a shear and the PSF is
given by
\begin{equation}\label{eq:KSBformula}
 \e -\e_\mathrm{s} = \Pg \gamma + P_\mathrm{sm} q^*;
\end{equation}
where $\e$ and $\e_\mathrm{s}$ are the observed and intrinsic ellipticities,
respectively, and
\begin{equation}\label{eq:Pg}
 \Pg = P_\mathrm{sh} - \left( \frac{P_\mathrm{sh}}{P_\mathrm{sm}}
 \right)^* P_\mathrm{sm} ,
\end{equation}
where the  tensors $P_\mathrm{sh}$ and $P_\mathrm{sm}$ can be
calculated from the galaxy light profile with the same filter scale of the
galaxy as for the  weight function in Eq.~(\ref{eq:Qij}); see
\cite{KSB}. The ratio $(P_\mathrm{sh}/P_\mathrm{sm})^*\equiv
(\mathrm{tr}P_\mathrm{sh}^*/\mathrm{tr}P_\mathrm{sm}^*)$ is calculated
from stars with the filter scale of the galaxy. The
second term in Eq.~(\ref{eq:Pg}) accounts for the circular smearing of
the isotropic part of the PSF. The stellar 
anisotropy kernel $q^*$ which is needed to correct for PSF anisotropy
can be calculated by noting that for stars $\e_\mathrm{s}^*=0$ and
$\gamma^*=0$,  so that
\begin{equation}\label{eq:ani}
  q^* = (P_\mathrm{sm}^*)^{-1} \e^*.
\end{equation}
The anisotropy-corrected ellipticity is then calculated by
\begin{equation}\label{eq:anicorr}
  \e^\mathrm{ani} =  \e - P_\mathrm{sm} q^*.
\end{equation}

In Fig.~\ref{fig:galani} we show for one galaxy field the uncorrected
ellipticities of the galaxies in the left panel and  the mean of the
anisotropy-corrected ellipticities over all galaxies in the  field
when correcting with different star field polynomials in the right
panel. There is a  shift in the negative $\e_1$  direction compared to
the mean uncorrected ellipticity, as is expected [see
Sect.~\ref{sc:PSF}]. The dispersion between different corrections is
much less than 1\% which shows that variations of the PSF anisotropy
are small.

\begin{figure}                             
 \resizebox{\hsize}{!}{\includegraphics{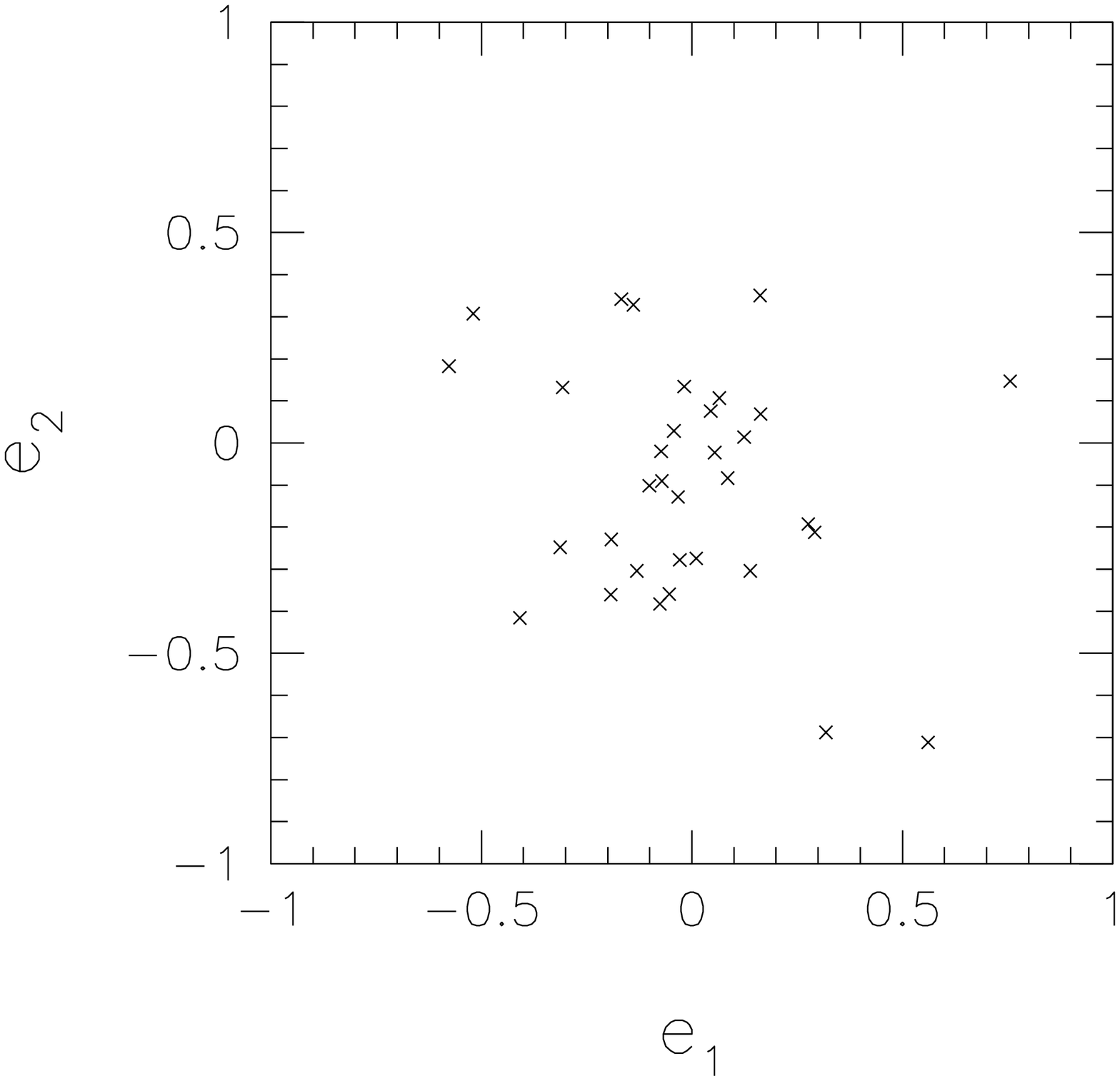}
                       \includegraphics{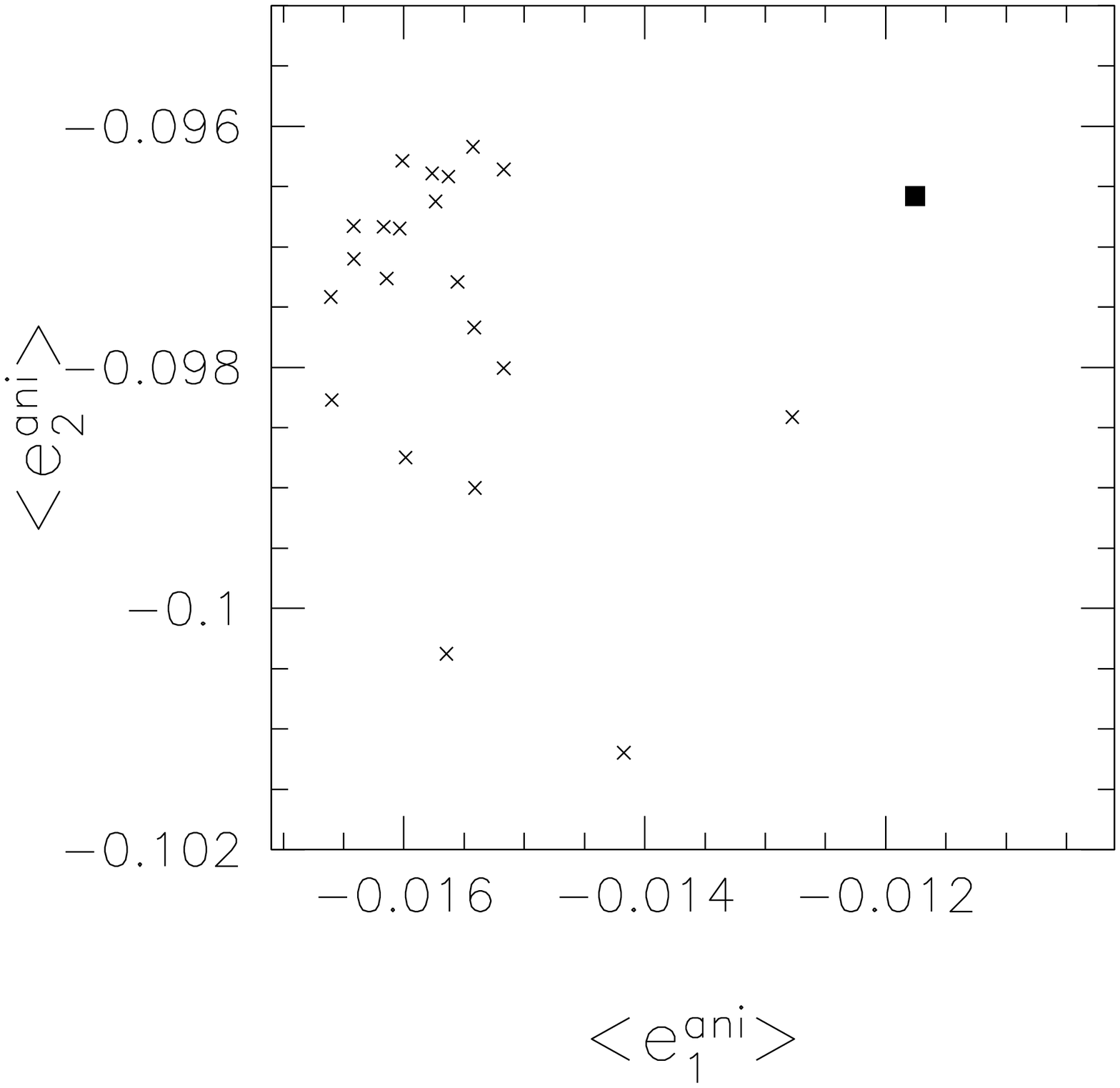}}
 \caption{For the galaxy field o46p02010\_3\_ass: in the left panel
 the uncorrected ellipticities of individual galaxies are shown, in 
 the right panel the mean of the anisotropy-corrected ellipticity
 $\ave{\e^\mathrm{ani}} $ over 
 the galaxy field for corrections with  different star fields;
 the square shows  the mean over the uncorrected ellipticities.  Note 
 the  different scalings of the panels. } 
 \label{fig:galani}
\end{figure}                               

In Fig.~\ref{fig:mgani} the mean ellipticity for all the galaxy fields is
shown with and without the PSF anisotropy correction.      It
illustrates that the PSF anisotropy correction changes the mean
ellipticity by an amount typically smaller than 1\%.  Also, the
dispersion between different PSF models from different star
fields is much less than 1\%, which means that the changes of the PSF
anisotropy seen in different star fields are sufficiently small to
allow us to use one (or a  suitable combination) of them for the
actual analysis.

The effect of PSF anisotropy corrections is considerably less than the
expected cosmic shear signal, which confirms our expectation that the
image quality of STIS is very well suited for our project.  We also
note that the mean PSF anisotropy correction points towards the
negative $\e_1$  direction, in agreement with the stellar ellipticity
plotted in Fig.~\ref{fig:aniexp}.

\begin{figure}                             
 \resizebox{\hsize}{!}{\includegraphics{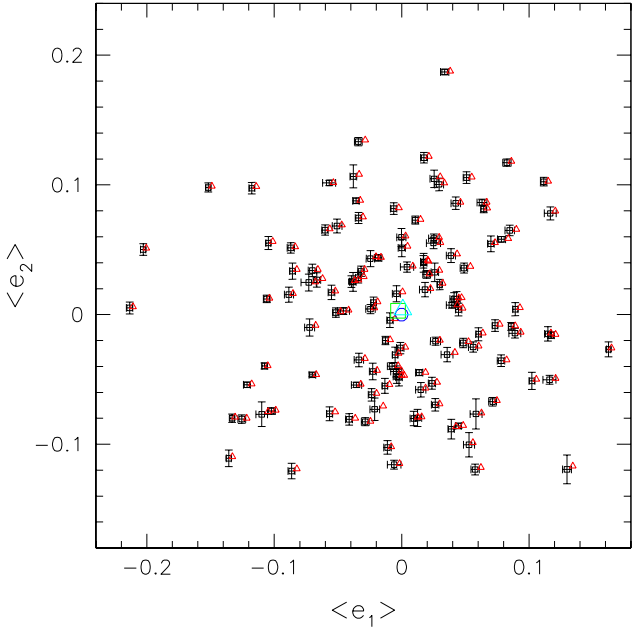}}
 \caption{For 121 galaxy fields we plot the mean uncorrected
 ellipticity of galaxies (triangles) as well as the mean anisotropy
 corrected ellipticity (squares). The error bars attached to the
 squares denote 3 times the dispersion of the field-averaged corrected
 ellipticities when the different PSF model fits are used; the error
 on the mean is much smaller than the symbols used. The shift of the
 corrected mean ellipticities towards negative $\e_1$ is expected from
 the behaviour of the stellar ellipticities plotted in
 Fig.~\ref{fig:aniexp}. The big triangle and big square in the centre
 denote the mean over all galaxy fields of the uncorrected and
 corrected mean ellipticities, respectively; the size of the symbols
 represent the 1$\sigma$ errors on the mean. The circle shows the
 origin for reference. The mean shear is compatible with zero.}
 \label{fig:mgani}
\end{figure}                               

\end{subsection}  

\begin{subsection}{Smearing correction}\label{ss:smearing}
The smearing corrected ellipticity of each galaxy is calculated by
\begin{equation}\label{eq:corr}
 \e^\mathrm{iso} = (\Pg)^{-1} \e^\mathrm{ani},
\end{equation}
see eqs.~(\ref{eq:KSBformula}) and~(\ref{eq:anicorr}), which is an
unbiased (provided that $\ave{\e_\mathrm{s}}=0$) but very noisy
estimate of the shear $\gamma$. We apply a scalar inversion of
$(\Pg)^{-1}= 2/\mathrm{tr} \Pg$ since it is less noisy than the full
tensor inversion, as found by \cite{E01}~(2001).  

To calculate $\Pg$ [see Eq.~(\ref{eq:Pg})] for each galaxy,
we need to estimate $(P_\mathrm{sh}/P_\mathrm{sm})^*$  from the light
profile of the stars with the filter scale of the galaxy [see
Eq.~(\ref{eq:Qij})]. We find that $(P_\mathrm{sh}/P_\mathrm{sm})^*$ is
spatially constant, as for most gound-based telescopes
(\cite{E01}~2001) and calculate the mean of this quantity over the stars in
a given field for different filter scales. We find that
$(P_\mathrm{sh}/P_\mathrm{sm})^*$ varies from star field to star field
(i.e., in time), as does the PSF anisotropy. If we take the mean over
the star fields for  different filter scales, we find that
$(P_\mathrm{sh}/P_\mathrm{sm})^*$  increases with filter scale and is
constant for large objects [see  Fig.~\ref{fig:pstar}]. This behaviour
is expected since small objects are typically more affected by
smearing of the PSF and therefore have a larger $P_\mathrm{sm}^*$.

\begin{figure}                             
 \resizebox{\hsize}{!}{\includegraphics{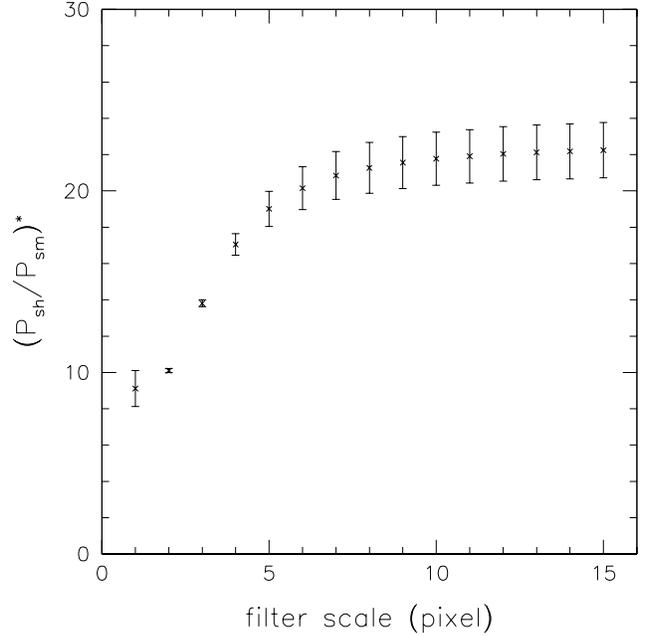}}
 \caption{The mean of $(P_\mathrm{sh}/P_\mathrm{sm})^*$ over all
 stars which were used for the anisotropy correction is shown for 
 different filter scales. The error bars show the dispersion between 
 different star fields. }
 \label{fig:pstar}
\end{figure}                               

The error bars on  $(P_\mathrm{sh}/P_\mathrm{sm})^*$ for different
filter scales are due to a variation in the size of the stars from
field to field, as is shown in Fig.~\ref{fig:rhstar}, where we plot
the mean half-light radius of all the stars in a star field  vs.\ the
time when the field 
was observed. The mean size of stars varies by about 0.2 subsampled STIS
pixels. We find that in fields with larger mean half-light radius,
$(P_\mathrm{sh}/P_\mathrm{sm})^*$ is bigger for all filter scales,
i.e., the smearing correction is larger.   The increasing error bars
for larger filter scale are due to the fact that stars are small
($\approx$2.3 subsampled STIS pixels) and therefore we detect more and
more (background) noise if we go to larger filter scales.

\begin{figure}                             
 \resizebox{\hsize}{!}{\includegraphics{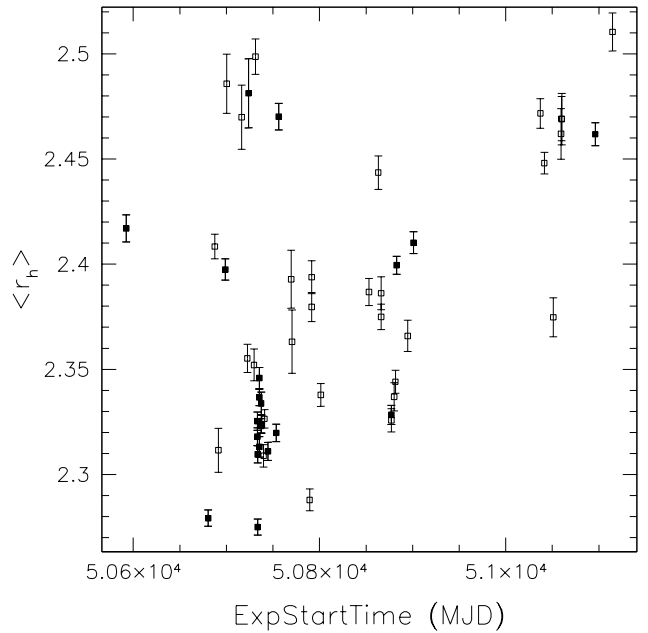}}
 \caption{The mean half-light radius of  all stars in a star field
 vs.\ the time when the  star field was observed. The filled squares 
 represent those star fields which were selected for the anisotropy 
 correction of the galaxy fields. }  \label{fig:rhstar}
\end{figure}                               

Since for the smearing correction one has to divide by $\Pg$ [see
Eq.~(\ref{eq:corr})],  objects with small values of $\Pg$ can get 
unphysically big ellipticities, which dominate the cosmic shear signal
even after introducing a weighting of galaxies as shown in the
following section. We therefore decide to introduce a cut in $\Pg$ in
addition to the weighting, requiring that `good' galaxies should have
$\sqrt{\det\Pg} \ge 0.2$ (in short: $\Pg>0.2$). The effect of
different cuts in $\Pg$ on the cosmic shear measurement  is discussed
in Sect.~\ref{ss:resw}. 

In Fig.~\ref{fig:pelli} we show the probability distribution of the
fully corrected ellipticities for all galaxies with $\Pg>0.2$ on all galaxy
fields. The mean  ellipticity is compatible with zero and the
dispersion in both components is 26\%. The dispersion is consistent with that
found by \cite{Hu98}~(1998), if 
one takes into account the different mean magnitude of the galaxies
and the different pixel scale. The kurtosis is consistent with
a Gaussian; the skewness is slightly larger than one
would expect given the number of galaxies. 

\begin{figure}                             
 \resizebox{\hsize}{!}{\includegraphics{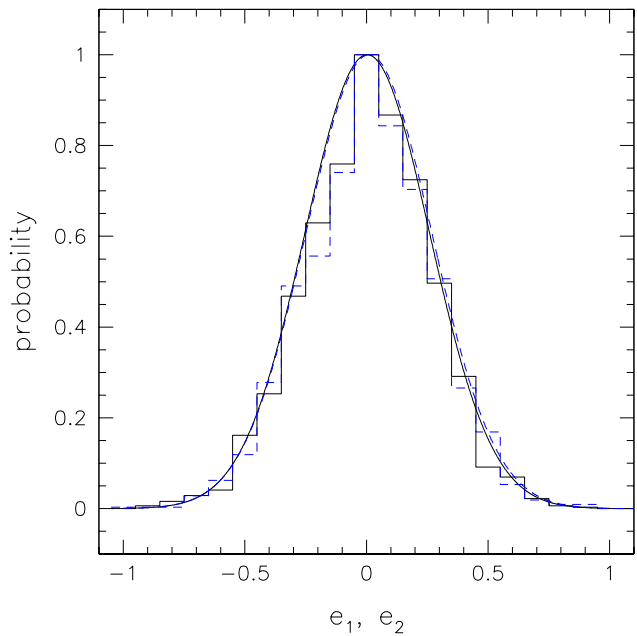}}
 \caption{Histogram of the probability distribution of the fully
 corrected ellipticities of all galaxies with $\Pg>0.2$ in all fields, $\e_1$
 solid,  $\e_2$ dashed. The curves show Gaussians with  mean and
 dispersion from the histograms. The mean ellipticity is very close
 to zero and $\sigma_1 = \sigma_2 = 26\%$.   } \label{fig:pelli}
\end{figure}                              

For the associations which were observed several times at
different visits, we compared the
catalogues of objects obtained from the different exposures and find
that the number of reliable detections ($\mathrm{S/N}>2$) differs. Comparing
the raw as well as the fully corrected ellipticities, we 
find that they agree very well regarding orientation angle and modulus
of the ellipticity, with the exception of a few faint extended
objects with low signal-to-noise ratio ($\mathrm{S/N}<10$). 

\end{subsection}

\begin{subsection}{Weighting scheme}{\label{ss:weight}}
The PSF correction we apply to the galaxy ellipticities amplifies the
errors in the measurement, which can
eventually produce unphysical ellipticities much larger than unity, as
noted in the previous section. 
For weak shear the ensemble of corrected ellipticities measured is the
best estimate one has of the intrinsic (unlensed) ellipticity
distribution. But this distribution is broadened and the 
high ellipticity tail is more pronounced because of the noise
amplification. 
With the weighting scheme we want to minimize the influence of objects
with high ellipticities which most probably originate from noise.  

Our expectation is that objects which are small and/or have low S/N ar
most affected by noise. For each galaxy we search for the 20 next
neighbours in a  two dimensional parameter space of 
half-light radius and signal-to-noise ratio (see \cite{E01}~2001)
containing all galaxies from all galaxy fields.  Since
the average number density on our galaxy fields is only 18, we have to
put all galaxies together in one catalogue to do the next neighbour
search. For the STIS images the galaxies have very similar
properties in the chosen parameter space, therefore combining them in
one catalogue is maintainable.   

Since the scalings in $\rh$ and S/N are very different we
transform the coordinate axes in the following way: we sort the
objects in ascending order in both coordinates separately and assign 
the running number as the new coordinate. The 20 next neighbours are then
found in this new parameter space.  
For each galaxy we calculate the ellipticity dispersion from the $M$
next neighbours
\begin{equation}
 \sigma_\mathrm{NN}^2 = \frac{1}{M-1}\sum_{i=1}^{M} \left( \e_i^2 -
 \ave{e}^2 \right) ,  
\end{equation}
which we assign as the measurement error to the galaxy. 
The weights for the galaxies are obtained by a simple  $w_\mathrm{NN}
= 1/\sigma_\mathrm{NN}^2$ weighting

The dispersion $\sigma_\mathrm{NN}$ and the corresponding weights
$w_\mathrm{NN}$ are shown in Fig.~\ref{fig:wnn}.   
The correlations seen in the plot are due to objects with similar properties,
e.g.\ at $\rh \approx 7$ there are several elliptical galaxies with a
compact, bright centre. 

\begin{figure}                             
 \resizebox{\hsize}{!}{\includegraphics{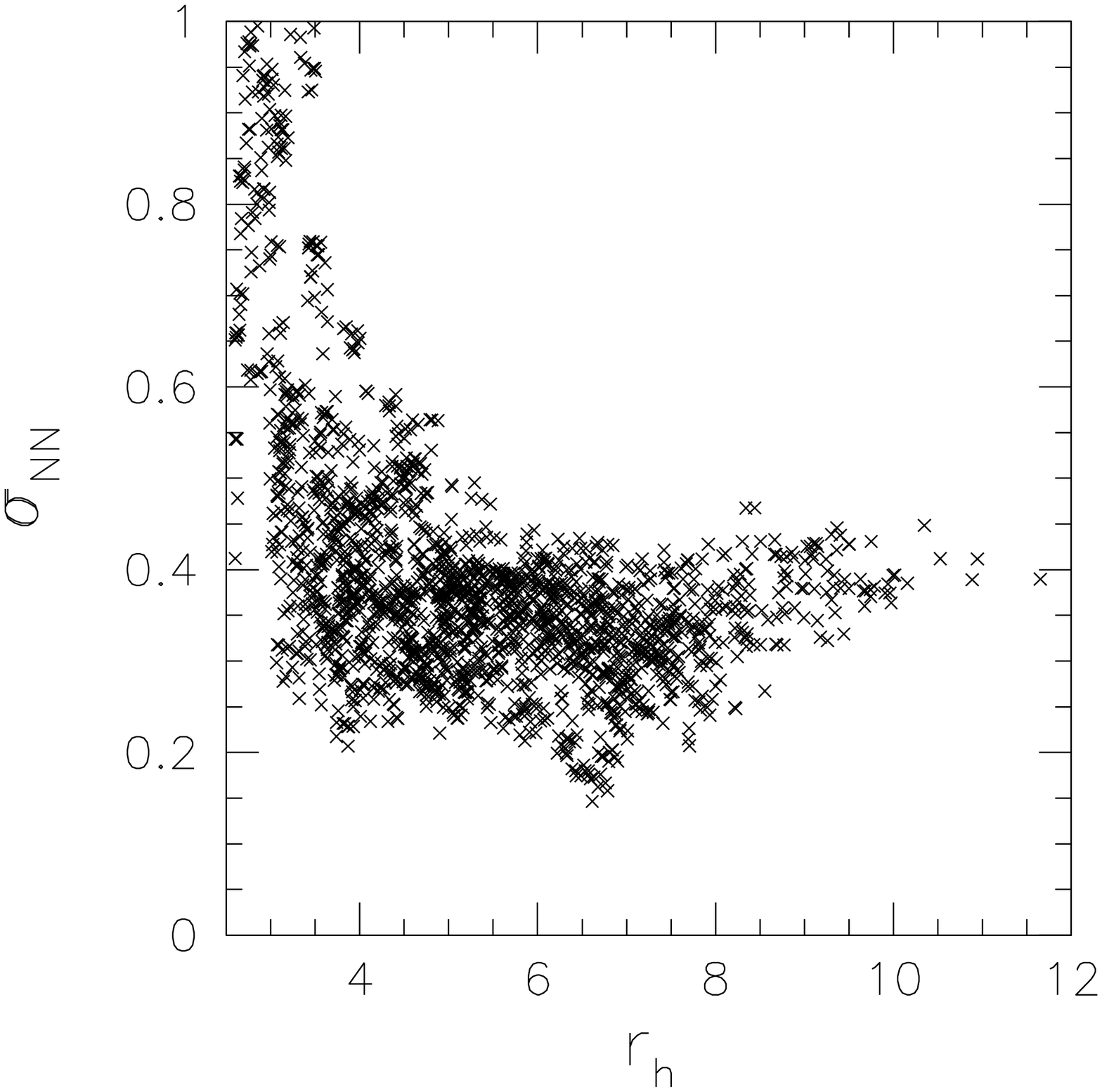}
                       \includegraphics{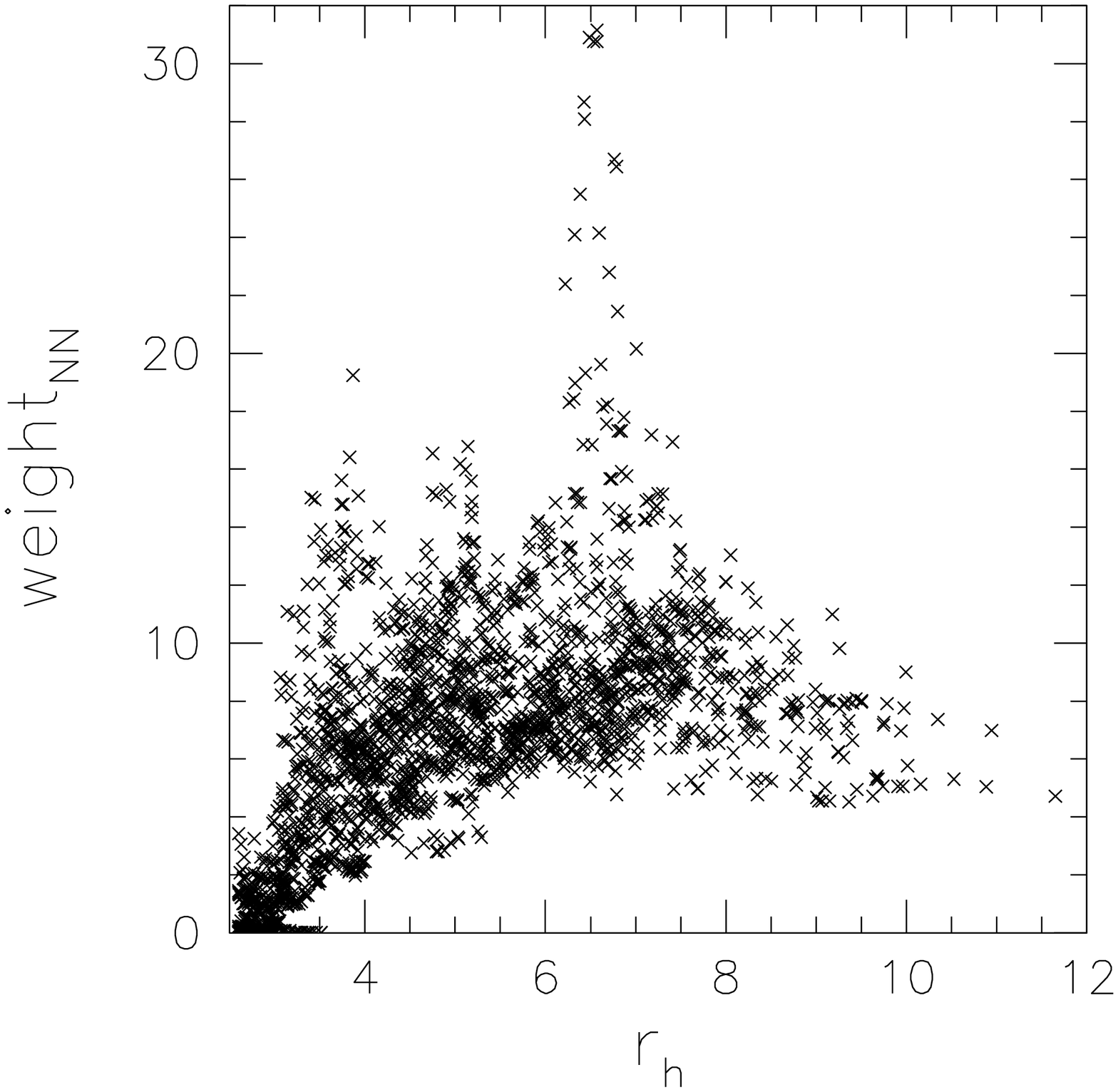}}
 \caption{In the left panel we show the dispersion 
 $\sigma_\mathrm{NN}$ of individual galaxy ellipticities for the 
 20 next neighbours in the $\rh$--S/N parameter space vs.
 half-light radius. In the right panel the weight deduced from this 
 dispersion is  shown.}
 \label{fig:wnn}
\end{figure}                               

\end{subsection}

\end{section}

\begin{section}{Results}\label{sc:res}

\begin{subsection}{Estimation of the cosmic shear}\label{ss:res}
The procedure to estimate the cosmic shear is the following:
\paragraph{PSF correction:}
The raw ellipticities of the galaxies are corrected for PSF anisotropy
with the polynomials from the 21 star fields. For each of the star
fields we also calculate $(P_\mathrm{sh}/P_\mathrm{sm})^*$ for
different filter scales and from that $\Pg$ for the filter
scale of the galaxy. For each galaxy we then have 21 anisotropy
corrected ellipticities and 21 smearing corrected tensors $\Pg$. The
mean correction is obtained by then averaging over the star
fields. From the mean anisotropy-corrected ellipticities and the mean
$\Pg$ for each galaxy we calculate the fully corrected ellipticity
[see Eq.~(\ref{eq:corr})].

\paragraph{Cosmic Shear:}
The fully corrected ellipticity is an unbiased (although very noisy)
estimate of the shear. Therefore, we use it to calculate the cosmic
shear estimator for each galaxy field, given in
eqs.~(\ref{eq:csn1}) and (\ref{eq:csn2}). By taking the average over all the
galaxy fields as in Eq.~(\ref{eq:cs2}) we obtain the measured cosmic
shear value in our data. The error on the cosmic shear result is
calculated as the statistical error on the mean, which for a weighted mean 
\begin{equation}
 \ave{x}= \frac{\sum_i w_i x_i}{ \sum_i w_i} 
\end{equation}
is given by
\begin{equation}\label{eq:sigxm}
 \sigma^2_{\ave{x}} = \frac{\sum_i w_i^2}{\sum_i w_i
 \left(\left(\sum_i w_i\right)^2 - \sum_i w_i^2\right)} \sum_i w_i
 (x_i - \ave{x})^2 .
\end{equation}
 
We also compute the confidence level of the results from the
dispersion of the probability distribution of the cosmic shear
estimator in  Eq.~(\ref{eq:cs2}) obtained  by randomizing  the galaxy
orientations [Fig.~\ref{fig:pdist}]. If we use all galaxy fields and
weight individual galaxies according to $w_\mathrm{NN}$ we
find in 2.9\% of the randomizations a value higher than the actually
measured one. 

\begin{figure}                  
 \resizebox{\hsize}{!}{\includegraphics{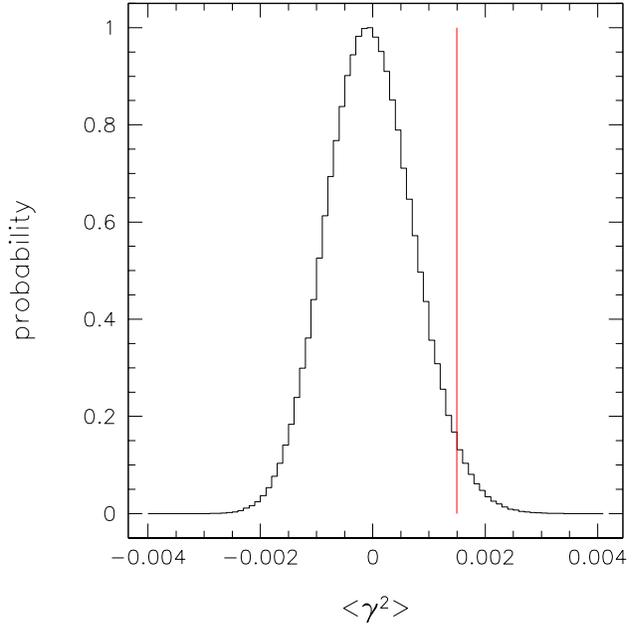}}
 \caption{Probability distribution of the cosmic shear estimator
 calculated from the data by randomizing the orientations of galaxies
 with $\Pg>0.2$ in all galaxy  fields. Individual galaxies are
 weighted with $w_\mathrm{NN}$ and the galaxy fields are weighted by
 the number of galaxies on each field. The
 vertical line indicates the actually measured value [see 
 Table~\ref{tab:res}; 1st row].} 
 \label{fig:pdist}                              
\end{figure}                               

Our results for the cosmic shear and the  corresponding errors [see
Eq.~(\ref{eq:sigxm})] are shown in Table~\ref{tab:res} for all
galaxy fields and for fields with more than 10 (15) galaxies per
field.  Note that in some of the  fields there are fewer than 10
galaxies per field (but still more than 5), since we restrict our 
analysis to ``good'' galaxies, i.e., objects with $\rh>2.6$ subsampled
STIS pixels [see Sect.~\ref{sc:cat}] and $\Pg \ge 0.2$ [see
Sect.~\ref{ss:smearing}]. 
Using only fields with a higher number density of galaxies we find a
larger cosmic shear signal, whereas the error stays
more or less the same although we average over fewer galaxy fields,
even if we do not apply any weighting to the galaxy fields. 

We also investigated the behaviour of the cosmic shear estimator in
bins of roughly equal number of fields for different number densities
of galaxies and find $\cs=(-8.1 \pm 26.5) \times 10^{-4}$ for fields
with $1< N\le10$, $\cs=(12.0 \pm 18.11) \times 10^{-4}$ for fields
with  $10< N\le16$ and $\cs=(23.4 \pm 14.0) \times 10^{-4}$ for fields
with $N>16$ when weighting individual galaxies with $w_\mathrm{NN}$
and galaxy fields with $W_\mathrm{f}=N$. This shows clearly that the
cosmic shear estimate increases if we use fields with higher number
density of galaxies. These fields have typically a larger exposure
time and therefore probably probe higher redshifts. The main source
for the better signal in these fields is the reduced error which is
due to a better sampling of the intrinsic ellipticity distribution.

\begin{table*}
\center
\caption{Results for the cosmic shear estimator and errors for
different  minimum number of galaxies per field. 
$N$ is the number of  galaxies per field, $N_\mathrm{f}$ is the number
of galaxy fields with $N\ge N_\mathrm{min}$.  
The first block shows the results for different cuts in $\Pg$, where
we weight individual galaxies with $w=w_\mathrm{NN}$ and the galaxy
fields with $W_\mathrm{f}=N$.
In the next block the results are given for weighting individual
galaxies or not. 
The last block shows the results for the cosmic shear if we apply
different weights to galaxy fields, weighting them equally
($W_\mathrm{f}=1$), weighting by the number of galaxies per field
($W_\mathrm{f}=N$) or by the square of the number of galaxies
($W_\mathrm{f}=N^2$).  
Note that we repeat the result with $\Pg>$0.2, $w=w_\mathrm{NN}$,
$W_\mathrm{f}=N$ in each block for easier comparison of the results. } 
\label{tab:res}
\begin{tabular}{|l|rrr|rrr|rrr|}
\hline 
 & \multicolumn{3}{c|}{all} & \multicolumn{3}{c|}{$N\ge10$} &
   \multicolumn{3}{c|}{$N\ge15$}\\ 
 & $N_\mathrm{f}$ & $\cs$ & $\sigcs$ & 
   $N_\mathrm{f}$ & $\cs$ & $\sigcs$ & 
   $N_\mathrm{f}$ & $\cs$ & $\sigcs$ \\
 && $\times 10^{4}$ & $\times 10^{4}$ && $\times 10^{4}$ & $\times 10^{4}$ && 
    $\times 10^{4}$ & $\times 10^{4}$  \\
\hline 
\hline 
\multicolumn{10}{|c|}{different cuts in $\Pg$; $w=w_\mathrm{NN}$; $W_\mathrm{f}=N$}\\
\hline 
$\Pg>$0.2 & 121 &14.96 &11.61 & 90 &16.90 &11.42 & 49 &26.33 &12.28   \\
$\Pg>$0.1 & 121 &20.14 &12.37 & 92 &21.39 &12.02 & 50 &28.49 &12.46   \\
$\Pg>$0.0 & 121 &20.41 &12.34 & 92 &21.49 &12.08 & 54 &31.96 &14.05   \\
\hline 			       		   	   
\hline 
\multicolumn{10}{|c|}{different weighting of individual galaxies; $\Pg>$0.2; $W_\mathrm{f}=N$}\\
\hline 
$w=$1              & 121 & 9.18 &10.17 &90 &10.29 &10.05 & 49 &19.03 &10.68  \\
$w=w_\mathrm{NN}$  & 121 &14.96 &11.61 &90 &16.90 &11.42 & 49 &26.33 &12.28  \\
\hline 							       
\hline 
\multicolumn{10}{|c|}{different weighting of galaxy fields; $\Pg>$0.2; $w=w_\mathrm{NN}$}\\
\hline 
$W_\mathrm{f}=1$    & 121 &10.68 &11.81 & 90 &14.73 &11.26 & 49 &31.29 &12.36  \\
$W_\mathrm{f}=N$    & 121 &14.96 &11.61 & 90 &16.90 &11.42 & 49 &26.33 &12.28  \\
$W_\mathrm{f}=N^2$  & 121 &15.32 &16.74 & 90 &15.77 &16.27 & 49 &19.61 &16.08  \\
\hline 							       
\end{tabular}
\end{table*}

\end{subsection}

\begin{subsection}{Effect of selection of galaxies and weighting}\label{ss:resw}
In the first three rows of Table~\ref{tab:res} we present the results for
the cosmic shear for different cuts in $\Pg$,
weighting individual galaxies with $w_\mathrm{NN}$. 
If we lower the cut in $\Pg$, including galaxies whose high
ellipticity is most probably only due to noise,  we find 
that the cosmic shear estimator and the error increase. 

In the second block of Table~\ref{tab:res} we compare the cosmic shear
estimator for weighting or not weighting individual galaxies. We find
that the signal increases by one half for all bins if we apply weighting,
whereas the error increases only slightly. 

In the last block of Table~\ref{tab:res} we give the values for our
cosmic shear estimator applying different weights to the galaxy
fields. We compare the results for applying no weighting at all,
weighting by $W_\mathrm{f}=N$, which is the optimum weight if the
error is only due to Poisson noise and weighting by
$W_\mathrm{f}=N^2$. We do not discuss a weighting by exposure time,
which would give deeper fields with a higher cosmic shear signal a
larger weight, since our errors are mainly due to Poisson noise. Of
course there is a strong correlation between exposure time and number
density and therefore the weighting by $N$ reflects the deepness of
the fields, but there is also the effect of clustering.  
If we increase the weight of fields with many galaxies per field, the 
cosmic shear estimator increases if we take all fields, increases and
then decreases if we take only fields with more than 10
galaxies and decreases if we take more than 15 galaxies per field. The
error on the cosmic shear is the same for the $W_\mathrm{f}=1$ and
$W_\mathrm{f}=N$ weighting and is much larger for the $W_\mathrm{f}=N^2$
weighting. This weighting obviously gives a too large weight to fields
with a very high number density of galaxies. 

\end{subsection}

\begin{subsection}{Effect of PSF corrections}\label{ss:rescorr}
In order to estimate how much the PSF corrections affect our cosmic
shear result we did the cosmic shear analysis without either
anisotropy or smearing correction or both. The corresponding results
are found in the first block of Table~\ref{tab:res2}. Without the
anisotropy correction but  including the smearing correction
($\e^\mathrm{raw}$,$\Pg$) the result does not change very much, which
confirms that the STIS PSF anisotropy is sufficiently small.  However,
if we leave out the smearing correction
($\e^\mathrm{ani}$,$P_\mathrm{sh}$ and
$\e^\mathrm{raw}$,$P_\mathrm{sh}$), we find that the  cosmic shear
estimator decreases by a factor of roughly three and also  the dispersion
is much smaller. This is not an effect of the different number of
galaxy fields included in the averaging. The cosmic shear estimate is
smaller because  smearing tends to make objects rounder,  and
therefore any quantity calculated from the ellipticities is
smaller. The dispersion decreases because $\Pg$ is much noisier than
$P_\mathrm{sh}$.
In the last two rows of the first block of Table~\ref{tab:res2}
we also give the results for the cosmic shear estimate if we apply
\emph{neither} PSF correction \emph{nor} weighting. This demonstrates
that the detection depends neither on the specific weighting scheme nor
on the \emph{necessary} smearing correction and can be seen in
uncorrected data. However, in order to make an unbiased estimate of the
shear dispersion, the PSF and smearing corrections must be taken into
account and our particular choice of the weighting scheme minimizes the
error on this estimate.

In the second block of Table~\ref{tab:res2} we show the cosmic shear
estimator if we 
apply the PSF corrections from the star fields individually. Note that the
number of galaxies per field (and therefore the number of fields with
$N\ge N_\mathrm{min}$) also changes when correcting with
different star fields, since the smearing correction
$(P_\mathrm{sh}/P_\mathrm{sm})^*$ is different from star field to star
field (as noted in Sect.~\ref{ss:smearing}).  Therefore, for some
galaxies $\Pg$ is above or below the cut of 0.2 we introduced to
reject objects with unphysically large ellipticities. Comparing the
various results for the cosmic shear measurement, we find 
$\cs =( 16.0 \pm 1.2 (\mathrm{PSF}) \pm 11.8 (\mathrm{stat}))\times 10^{-4}$ 
for all fields, which shows that the difference between the star field
corrections is much smaller than the statistical error.

\begin{table*}
\center
\caption{Results for the cosmic shear for different PSF corrections. 
The first block shows the results if we do not correct for PSF effects:
the first Col. indicates if we use raw ellipticities
($\e^\mathrm{raw}$) or anisotropy corrected ellipticities
($\e^\mathrm{ani}$) and if we apply  the smearing correction ($\Pg$)
or not ($P_\mathrm{sh}$).  The first row gives the fully corrected
result [see Table~\ref{tab:res}] for reference. 
The next block shows the results when we apply PSF corrections from
the individual star fields. 
The results quoted were obtained weighting individual galaxies
[see Eq.~(\ref{eq:csn2})] with $w=w_\mathrm{NN}$, requiring
$\Pg>0.2$ (or $P_\mathrm{sh}>0.2$), and weighting the galaxy fields by
$W_\mathrm{f}=N$ unless otherwise indicated. }
\label{tab:res2}
\begin{tabular}{|l|rrr|rrr|rrr|}
\hline 
 \multicolumn{1}{|c|}{$e$, $P$} & \multicolumn{3}{c|}{all} & 
  \multicolumn{3}{c|}{$N\ge10$} & \multicolumn{3}{c|}{$N\ge15$}\\ 
 \multicolumn{1}{|c|}{or}       & $N_\mathrm{f}$ & $\cs$ & $\sigcs$ &
  $N_\mathrm{f}$ & $\cs$ & $\sigcs$ & $N_\mathrm{f}$ & $\cs$ & $\sigcs$ \\
 \multicolumn{1}{|c|}{starfield} && $\times 10^{4}$ & $\times 10^{4}$
  && $\times 10^{4}$ & $\times 10^{4}$ && $\times 10^{4}$ & $\times 10^{4}$ \\
\hline 
\hline 
$\e^\mathrm{ani}$, $\Pg$           & 121 &14.96 &11.61 &  90 &16.90 &11.42 & 49 &26.33 &12.28  \\
$\e^\mathrm{raw}$, $\Pg$           & 121 &16.14 &11.73 &  90 &18.25 &11.52 & 49 &27.61 &12.43  \\
$\e^\mathrm{ani}$, $P_\mathrm{sh}$ & 121 & 5.27 & 6.00 & 101 & 5.33 & 6.04 & 67 &10.33 & 6.58  \\
$\e^\mathrm{raw}$, $P_\mathrm{sh}$ & 121 & 5.69 & 6.02 & 101 & 5.75 & 6.06 & 67 &10.87 & 6.58  \\
$\e^\mathrm{raw}$, $P_\mathrm{sh}$; $w=1$; $W_\mathrm{f}=N$ & 121 & 4.63 & 4.49 &  101 & 4.80 & 4.50 &  67 & 8.72 & 4.68 \\
$\e^\mathrm{raw}$, $P_\mathrm{sh}$; $w=1$; $W_\mathrm{f}=1$ & 121 & 2.49 & 4.39 &  101 & 2.75 & 4.40 &  67 & 9.84 & 4.60 \\
\hline 
o3zf01\_3 & 121 &15.23 &12.24 & 90 &17.23 &12.00 & 49 &26.34 &13.05   \\
o46p1u\_1 & 121 &17.58 &11.44 & 90 &19.97 &11.54 & 50 &26.69 &11.93   \\
o46p4p\_3 & 121 &14.61 &11.98 & 89 &16.90 &11.77 & 48 &23.67 &12.74   \\
o48b3b\_1 & 121 &18.29 &13.60 & 88 &21.73 &13.11 & 49 &29.90 &14.15   \\
o48b3w\_3 & 121 &15.61 &11.57 & 90 &17.57 &11.41 & 50 &27.10 &12.13   \\
o48b3x\_3 & 121 &15.64 &11.52 & 90 &17.70 &11.41 & 50 &27.11 &12.22   \\
o48b41\_3 & 121 &15.93 &11.50 & 90 &18.01 &11.40 & 50 &26.98 &12.12   \\
o48b42\_1 & 121 &15.33 &11.26 & 91 &17.31 &11.04 & 50 &25.71 &11.76   \\
o48b45\_3 & 121 &16.02 &11.51 & 90 &18.15 &11.39 & 50 &27.49 &12.21   \\
o48b4j\_3 & 121 &15.35 &11.38 & 90 &17.37 &11.27 & 50 &26.67 &12.06   \\
o48b4k\_3 & 121 &15.79 &11.50 & 90 &17.93 &11.41 & 50 &27.36 &12.26   \\
o48b4v\_1 & 121 &18.07 &11.52 & 91 &20.55 &11.57 & 50 &27.04 &12.13   \\
o48b4w\_1 & 121 &15.36 &11.28 & 90 &17.38 &11.17 & 50 &25.98 &11.86   \\
o48b5j\_3 & 121 &15.30 &11.38 & 90 &17.28 &11.25 & 50 &26.69 &12.04   \\
o48b74\_1 & 121 &15.77 &11.38 & 90 &17.84 &11.32 & 50 &25.84 &11.93   \\
o48b8t\_1 & 121 &17.68 &13.03 & 87 &21.44 &12.62 & 49 &28.24 &13.61   \\
o48b9g\_2 & 121 &15.15 &11.35 & 90 &17.19 &11.23 & 50 &26.55 &12.04   \\
o4k1ez\_1 & 121 &14.76 &12.10 & 89 &16.95 &11.84 & 48 &23.85 &12.74   \\
o4lycf\_1 & 121 &15.05 &11.40 & 90 &16.99 &11.26 & 50 &26.31 &12.02   \\
o4lydt\_1 & 121 &16.20 &12.39 & 90 &18.27 &12.15 & 49 &27.94 &13.12   \\
o4xcll\_1 & 121 &18.21 &12.97 & 87 &21.93 &12.52 & 49 &28.43 &13.45   \\
\hline 
\end{tabular}
\end{table*}

In Fig.~\ref{fig:csind} we plot the mean cosmic shear estimators of
the galaxy fields when correcting with different star fields
vs.\ number of galaxies per field.  As noted above, due to the cut in
$\Pg$ the number of galaxies per field can differ.  The very large
vertical error bars are  
due to different numbers of galaxies per field (including galaxies
with high ellipticities or not). However, if we remove this effect by
applying the  cut on the mean $\Pg$ over all star fields rather than
the $\Pg$ obtained from correcting with an individual star field, we still
find variations of the cosmic shear estimator per galaxy field up to
$\Delta \csn = 0.004$, which is correlated with the mean shear per
field. This can be understood by noting that 
\begin{equation}
 \csn := \frac{1}{N_n (N_n-1)} \, \sum_{i\neq j} \e_{in} \e_{jn}^\star =
         \ave{\gamma_n}^2 - \frac{1}{N_n-1} \, \sigma_\mathrm{s}^2 ;
\end{equation}
then $\Delta\csn = 2 \ave{\gamma_n} \Delta \ave{\gamma_n}$. 
The spread in the cosmic shear values obtained from different fields
decreases for larger $N$, which shows that fields with more galaxies
are less affected by noise. Note that the cosmic shear estimator can
also be negative. 

\begin{figure}                             
 \resizebox{\hsize}{!}{\includegraphics{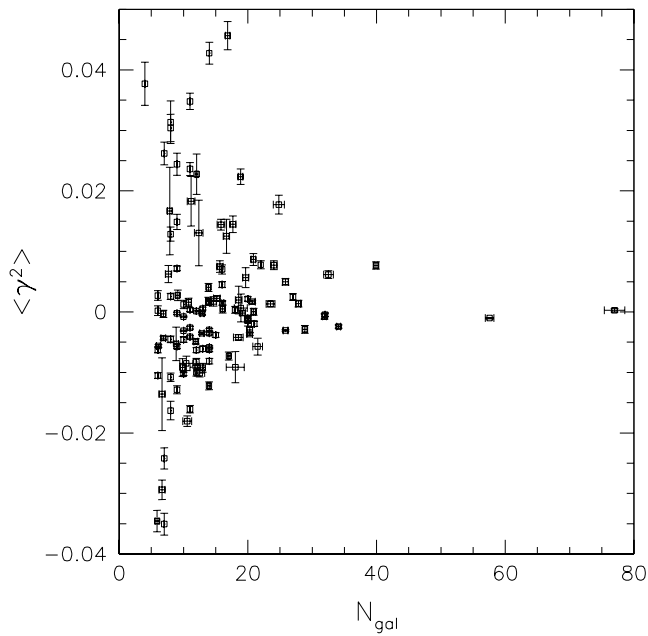}}
 \caption{For each of the  121 galaxy fields the cosmic shear was
 measured from ellipticities corrected with 21 individual star
 fields. We show the mean of the cosmic shear estimator over the star
 fields as boxes vs.\ the number of  galaxies in the field. The error
 bars show the 1$\sigma$ dispersion of individual values from the
 mean. Note that the number of galaxies per field can change because
 of the cut in $\Pg$. } \label{fig:csind}
\end{figure}                               

\end{subsection}

\begin{subsection}{Simulations}\label{ss:sim}
In order to test the analysis pipeline we produced 140 uncorrelated
galaxy fields and 20 star fields ($1\arcmin \times 1\arcmin$). 

Catalogues with intrinsic properties for the galaxies were produced with
the \texttt{Stuff}\footnote{Part of the Terapix software suite available at
\texttt{http://terapix.iap.fr/soft/}} program (see \cite{E01}~2001). The
galaxies were taken 
to be at the same redshift ($z=1)$.  The effect of cosmic shear
was simulated by doing ray tracing through a number of lens planes
which were produced from N-body simulations in an open CDM cosmology 
(\cite{JSW00}~2000). Maps of the shear field produced from these
simulations were sampled with randomly placed fields like the STIS fields. 
The cosmic shear signal on arcminute scales in these fields 
is $\sqrt{\cs}=2.6\%$ (\cite{JSW00}~2000).

From the sheared galaxy catalogues, images were produced with
\texttt{Skymaker}\footnote{Part of the Terapix software suite available at
\texttt{http://terapix.iap.fr/soft/}} (see \cite{E01}~2001), where the
parameters were tuned to 
match as closely as possible the physical characteristics of the STIS
CCD for exposure times of 400 s, so noise properties of individual
simulated images are very close to the ones observed in real
data. Star fields were created directly using \texttt{Skymaker} in the
stellar field mode. We focused on obtaining stars of similar
half-light radii as the observed ones, by setting the tracking error
type parameter in \texttt{Skymaker} to JITTER and its size to
0$\arcsecf$01. We simulated a round PSF and did not try to
include the effects of optical distortion of HST, since the simulations
were just done to test the analysis pipeline and its capability to
retrieve the input cosmic shear signal.  To simulate a coadded
exposure of 1600 s, we first created 4 individual 400 s images from the
same original sheared galaxy catalogue, but with positions randomly
dithered in order to process them in the same way as the real
data. Dithers could vary between 0 and 30 pixels. The pipeline was
able to recover the dithering within one tenth of a pixel as stated in
\cite{PaperI}.  The images were then drizzled and median averaged in
the same way as the archive data.

The analysis of the simulated fields was carried out following  the
same procedure for catalogue production and cosmic shear estimation as
described in Sect.~\ref{sc:cat} and~\ref{ss:res}. In particular, we used
the same parameters for simulated and real data. We find a cosmic
shear signal of $\sqrt{\cs}=2.3\%$ with $1.6\sigma$ significance ($\cs
= (5.1 \pm 3.2)\times10^{-4}$), which is slightly smaller than the input
value of 2.6\%.   Since we under- rather than
over-estimate the input signal we conclude that our procedure does not
introduce a spurious cosmic shear signal.

For a quantitative analysis, we have to generate more realistic
simulations which represent the  properties of our data better with
respect to different exposure times and therefore different redshift
distributions, in order to better understand their contribution to
the  cosmic shear signal. 
\end{subsection}

\end{section}

\begin{section}{Discussion}\label{sc:diss}
We report on our initial measurement of cosmic shear on scales below
one arcminute with STIS parallel archive data. 
Since any PSF anisotropy can mimic shear, we investigated the PSF
anisotropy of the STIS CCD in detail. 
Although the STIS PSF is not stable in time, we show that the
anisotropy on the STIS field is sufficiently small (on the order of
1\%) to carry out the cosmic shear analysis. It  changes the mean
shear on galaxy fields by much less than 1\% and does not affect our
estimate of the cosmic shear rms by more than 10\%.   
The smearing introduced by the PSF due to the small galaxy sizes at
faint magnitudes is much more important than the PSF anisotropy and
produces the biggest dispersion in our results. This is a rather
fundamental problem in that the galaxies are small and only NGST will
have a higher spatial resolution.  
 
To test our procedure for catalogue production and cosmic shear
estimation, we carried out simulations including the drizzling
procedure. The cosmic shear result we obtained from the
simulations is only slightly smaller than the input value, which shows that
our procedure does not introduce a spurious cosmic shear signal and
that our careful galaxy selection leads to a conservative estimation
of the cosmic shear. 

As discussed in Sect.~\ref{ss:resw} weighting of
individual galaxies (according to the measurement error) or weighting
of  galaxy fields (to account for Poisson noise) changes the result
substantially.  
If we restrict our analysis to fields with a higher number density of
galaxies, we find that the signal increases on average, although not
significantly. This would agree
with a cosmological interpretation of the signal: fields with a higher
number density of objects typically have a larger exposure time,
therefore they typically probe higher redshifts and one expects a
higher cosmic shear signal. 
For a quantitative interpretation of the cosmic shear
measurement an optimal weighting scheme still has to be found, which
will include redshift information on the STIS fields. 

Although we obtained a detection of the cosmic shear the error
bars are still large. As can be seen in eqs.~(\ref{eq:sigintr})
and~(\ref{eq:sigcv}) the errors in the data depend on both the number of
galaxies per field and on the number of fields. It is therefore
important to get more fields with higher number densities of
objects. The parallel  observations with STIS are currently being continued
in the frame of a GO cycle 9 parallel proposal (8562+9248, PI: P.~Schneider).

In Fig.~\ref{fig:cmpcs} we compare our result of
$3.87^{+1.29}_{-2.04}\%$ for the cosmic shear to 
the ones obtained by  other groups on larger  scales and to the
theoretically expected values when using different cosmological models
with a mean redshift of the sources of $\ave{z_\mathrm{s}}\approx
1.2$, which is appropriate for the groundbased measurements. 
With the STIS data we are probably probing at higher mean redshifts, but
since our observations with STIS were taken in the CLEAR filter mode,
we do not have much information about the redshift range of our
galaxies. Moreover, our fields have a large spread in exposure times
and therefore we effectively average over different cosmic shear
values.  
Multicolour (optical and near-IR) observations from the ground are
presently being
carried out to determine the redshift distribution of the galaxies in
the STIS fields, using photometric redshifts on the actual observed
STIS fields and their vicinity. With these
data, we will be able to interpret the cosmic shear measurement with
respect to other scales. This will be done in a forthcoming paper. 

\begin{figure}                  
 \resizebox{\hsize}{!}{\includegraphics{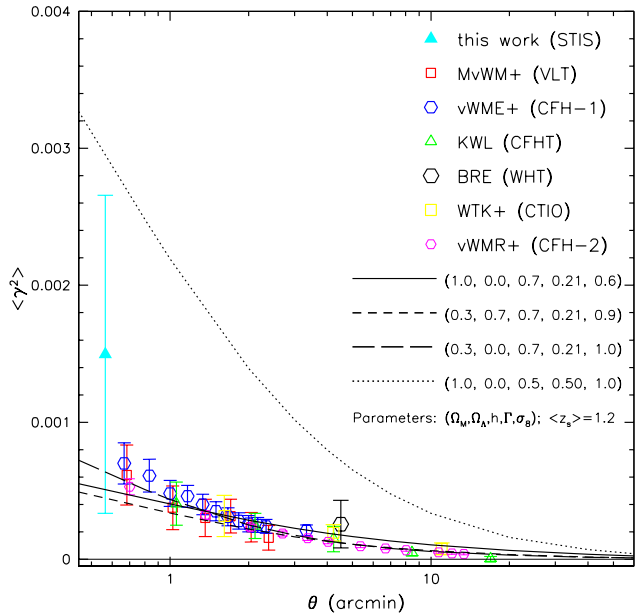}}
 \caption{Comparison of our cosmic shear result if we use all 
 galaxy fields (weighting individual galaxies by $w_\mathrm{NN}$ and
 galaxy fields with $N$) with
 measurements at larger angular scales from other groups and with
 model  predictions. 
 $\theta$ is the radius in arcmin at which the results were obtained;
 the cosmic shear results which were measured on a square field
 (STIS, BRE, KWL) are plotted at the radius of a circular field with the
 same area, the result from WTK is plotted at half the separation angle of
 the galaxies.
 The lines show the theoretical predictions if one uses
 different cosmological models, which are characterized by $\Omega_\mathrm{m}$,
 $\Omega_\Lambda$, $h$, $\Gamma$, and $\sigma_8$. The redshift
 distribution is taken from \cite{BBS}~(1996), with a mean source redshift
 of $\ave{z_\mathrm{s}}=1.2$. }  \label{fig:cmpcs}
\end{figure}                               

\end{section}

\begin{acknowledgements}
We thank Y. Mellier and L. van Waerbeke for fruitful discussions.
We thank L. King and J. Sanner for careful reading of the manuscript. 
This work was supported by the TMR Network ``Gravitational Lensing: New
Constraints on Cosmology and the Distribution of Dark Matter'' of the
EC under contract No. ERBFMRX-CT97-0172  and by the German
Ministry for Science and Education (BMBF) through the DLR under the
project 50 OR106.
RAEF is affiliated to the Space Science Department of the European
Space Agency. BJ acknowledges funding support from STScI/NASA. 
\end{acknowledgements}

\end{document}